\documentclass[journal]{IEEEtran}

\usepackage{epsfig} % for postscript graphics files
\usepackage{epstopdf}
\usepackage[dvipsnames]{xcolor} % for postscript graphics files
\usepackage{amsmath}
\usepackage{amsthm} % assumes amsmath package installed
\usepackage{amssymb}  % assumes amsmath package installed
\usepackage{subfigure}
\usepackage{soul}

\newtheorem{theorem}{Theorem}

\usepackage{enumerate}
\newtheorem{proposition}{Proposition}
\newtheorem{lemma}{Lemma}

\newtheorem{remm}{Remark}
\newenvironment{remark}{\begin{remm}\rm }{\hfill \hspace*{1pt} \hfill $\circ$\end{remm}}

\newcommand\hspa{\hspace{-.5cm}}
\newcommand\Ltwo{\ensuremath{{\mathcal L}_2}}
\newcommand\real{\ensuremath{{\mathbb R}}}

\DeclareMathOperator{\sat}{sat}
\DeclareMathOperator{\dz}{dz}
\DeclareMathOperator{\He}{He}

\def\trace{\mathrm{trace}}
\def\Q{\mathbb{Q}}
\def\S{\mathcal{S}}
\def\B{\mathrm{B}}

\newcommand{\ped}[1]{{_{\mathrm{#1}}}}

\def\tRO{\theta\ped{RO}}
\def\tSO{\theta\ped{SO}}
\def\tROC{\theta\ped{RwC}}
\def\tSOC{\theta\ped{SwC}}
\def\tSSOC{\theta\ped{SSwC}}
\def\V{\mbox{\rm Viol}}
\def\R{\mbox{\rm Rel}}

\begin{document}
% Do not put math or special symbols in the title.
\title{Robust linear static anti-windup\\ with probabilistic certificates}

\author{Simone~Formentin,~\IEEEmembership{Member,~IEEE,}
        Fabrizio~Dabbene,~\IEEEmembership{Senior~Member,~IEEE,}
		 Roberto~Tempo,~\IEEEmembership{Fellow,~IEEE,}
		 Luca~Zaccarian,~\IEEEmembership{Fellow,~IEEE,}
        and~Sergio~M.~Savaresi,~\IEEEmembership{Senior~Member,~IEEE}% <-this % stops a space
\thanks{Simone Formentin and Sergio M. Savaresi are with Dipartimento di Elettronica, Informazione e Bioingegneria, Politecnico di Milano, Piazza Leonardo da Vinci 32, 20133 Milano, Italy.}
\thanks{Fabrizio Dabbene and Roberto Tempo are with CNR-IEIIT, Corso Duca degli Abruzzi 123, Torino, Italy.}
\thanks{Luca Zaccarian is with CNRS, LAAS, 7 avenue du colonel Roche, F-31400 Toulouse, France,
Univ. de Toulouse, LAAS, F-31400 Toulouse, France, and Dip. di Ingegneria Industriale, University of Trento, Italy.}
\thanks{E-mail to: {\tt\small simone.formentin at polimi.it, fabrizio.dabbene at ieiit.cnr.it, roberto.tempo at polito.it, zaccarian at laas.fr, sergio.savaresi at polimi.it}.}
%\thanks{Manuscript received April ??, 2015; reised December ??, 2014.}
}

% % The paper headers
%\markboth{IEEE Transactions on Automatic Control,~Vol.~X, No.~X, XX~2015}%
%{Formentin \MakeLowercase{\textit{et al.}}: Robust anti-windup design}

\maketitle

%%%%%%%%%%%%%%%%%%%%%%%%%%%%%%%%%%%%%%%%%%%%%%%%%%%%%%%%%%%%
\begin{abstract}
In this paper, we address robust static anti-windup compensator design and performance analysis for saturated linear closed loops in the presence of nonlinear probabilistic parameter uncertainties via randomized techniques. The proposed static anti-windup analysis and robust performance synthesis  correspond to several optimization goals, ranging from minimization of the nonlinear input/output gain to maximization of the stability region or maximization of the domain of attraction. We also introduce a novel paradigm accounting for uncertainties in the energy of the disturbance inputs. 

Due to the special structure of linear static anti-windup design, 
wherein the design variables are decoupled from the Lyapunov certificates, 
we introduce a significant extension, called \textit{scenario with certificates} (SwC), of the so-called scenario approach for uncertain optimization problems. This extension is of independent interest for similar robust synthesis problems involving parameter-dependent Lyapunov functions. We demonstrate that the scenario with certificates robust design formulation is appealing because it provides a way to implicitly design the parameter-dependent Lyapunov functions and to remove restrictive assumptions about convexity with respect to the uncertain parameters. Subsequently, to reduce the computational cost, we present a sequential randomized algorithm for iteratively solving this problem.
The obtained results are illustrated by numerical examples.
\end{abstract}


\IEEEpeerreviewmaketitle

%%%%%%%%%%%%%%%%%%%%%%%%%%%%%%%%%%%%%%%%%%%%%%%%%%%%%%%%%%%%

\section{Introduction}

\IEEEPARstart{A}{nti-windup} designs correspond to control systems augmentations in light of actuator saturations, to mitigate the negative effects of the input nonlinearity. Their development has a history dating back to the era of analog controllers, more than half a century ago, and the most effective techniques are well illustrated in \cite{TarBook11,ZackAWbook11,TarbouriechIET09,ZackEJC09}. When robustness to parameter uncertainties must be taken into account, only recent results on suitable anti-windup constructions become available, all of them formulated in the deterministic robust control context. Some relevant examples correspond to \cite{MarcosTAC07,TurnerTAC07,Sofrony07,4380501,Post07,GaleaniAuto06,Forni10LPV,GrimmIJRNC04,ValmorbidaTAC13}, where several successful solutions differing in nature and architecture have been proposed.
While these available robust anti-windup solutions arise from different approaches and paradigms to the robust anti-windup problem, they mostly share the common feature of arising from a deterministic approach wherein a constant but unknown parameter belongs to a (known) compact set. Then, by suitable relaxations of the assumptions at the price of increased conservativeness, these sets are convexified to obtain numerically tractable approaches to the analysis and design problem.
In this paper we follow a radically different paradigm, arising from randomized methods for performance analysis and control design.

Randomized and probabilistic methods for control received a growing attention in the systems and control community in recent years \cite{TeCaDa:13}. These methods deal with the design of controllers for systems affected by possibly nonlinear, structured and unstructured uncertainties. One of the key features of these methods is to break the curse of dimensionality, \textit{i.e.}, uncertainty is ``lifted'' and the resulting controller satisfies a given performance for ``almost'' all uncertainty realizations. In other words, in this framework, we accept a ``small'' risk of performance violation.

One of the successful methods that have been developed in the area of randomized and probabilistic methods is the so-called scenario approach, which provides an effective tool for solving control problems formulated in terms of robust optimization \cite{CalCam:06tac}. In this case, the sample complexity, which is the number of random samples that should be drawn according to a given probabilistic distribution, is derived a priori, and it depends only on the number of design parameters $n_{\theta}$, and probabilistic parameters
called accuracy $\epsilon$ and confidence $\delta$. 

In parallel with these methods, sequential-based approaches have been developed, see for instance the recent sequential probabilistic validation techniques proposed in \cite{ATLR:15} and references therein.
In particular, in \cite{CDTV:15} an algorithm is proposed which, at each iteration, constructs a candidate controller, whose performance is then validated through a Monte Carlo approach. If the controller does not enjoy the required probabilistic performance specification, a new controller is designed based on new sample extractions. At each step of the sequence, a reduced-size scenario problem is solved. This method is usually effective in practical applications, even if its sample complexity cannot be determined a priori.
These methods may be used in specific control problems such as designing a common quadratic Lyapunov function. In these cases, however, the fact that a single common Lyapunov function should hold for all possible uncertainties leads to an overly conservative design. The same drawback is known in classical robust control, where the design of a common quadratic Lyapunov function requires an exponential number of computations \cite{ATRC:08,CalDab:08b}. For these reasons, parameterized Lyapunov functions have been developed and used in many robust control problems subject to uncertainty \cite[Chapter 19.4]{Barmish:94}.

Within the above surveyed context, the contribution of this paper is two-fold.
In the first part of the paper (Section~\ref{sec:SwC_big}), we develop a new framework, denoted as scenario with certificates, which is very effective in dealing with parameter-dependent Lyapunov functions. This framework continues the research originally proposed in \cite{Oishi:07a} for feasibility problems in the context of randomized methods. The main idea in this approach is to distinguish between \textit{design} variables $\theta$ and \textit{certificates} $\xi$ and has the advantage, compared to classical robust methods, that no explicit parameterization (linear or nonlinear) of the Lyapunov functions is required. In other words, the method is based on a ``hidden" parameterization of the Lyapunov functions, and has the clear advantage to reduce the conservatism compared to the methods based on the design of common Lyapunov functions.

In the second part of the paper (Sections~\ref{ref:aw} and \ref{sec:example}), we show the application of the scenario with certificates approach to anti-windup design and analysis in the presence of time-invariant uncertainty.
In particular, we concentrate on a specific anti-windup scheme for linear saturated plant-controller feedbacks: static direct linear anti-windup design (see, e.g., \cite[Part II]{ZackAWbook11}). Direct linear anti-windup corresponds to augmenting a linear saturated control design with a linear gain $D_{aw}$ driven by the excess of saturation $\dz(u) = u-\sat(u)$ and injecting suitable correction ``anti-windup'' signals at the state and output equation of the pre-designed windup-prone linear controller.
Several different performance optimization tasks are considered, and we present different alternatives in the subsections of Section~\ref{ref:aw}. A notable one, which is novel to  the anti-windup field and arises naturally from the proposed probabilistic context, is the one (in Section~\ref{sec:ui}) where the design minimizes an upper bound of the area spanned by the nonlinear \Ltwo\ gain curve, accounting for uncertain (but probabilistically known) energy of the external disturbance acting on the saturated closed loop. Each proposed performance metric is shown together with a robust performance analysis result that is not limited to the anti-windup context but is applicable to any uncertain linear closed loop subject to saturation in the classical LFT form. In all the above contexts, we will show that the probabilistic approach allows to reduce conservatism as well as to cope with uncertainty entering nonlinearly in the problem description, without overbounding it. The latter case has instead already been treated in various examples in the literature, where the trade-off between the robust and deterministic approach is usually referred to as \textit{probability degradation function}, see e.g. \cite[Ex. 11.1 and 12.1]{TeCaDa:13}.

Preliminary results in the direction of this paper were presented in \cite{FormentinCDC13,FormentinCDC14}.
In particular, in \cite{FormentinCDC13} the results were based on the classical scenario optimization approach.
In that formulation, both the certificates and the design variables were treated as optimization variables over the whole operating region, thus leading to a conservative solution, and - sometimes - to infeasibility. 
The scenario with certificates solution proposed here was then introduced in \cite{FormentinCDC14}, where we also provided preliminary results on the design of anti-windup compensators minimizing the nonlinear \Ltwo\ gain. 

As compared to these preliminary results, in this paper we fully exploit the potential of the proposed randomized approach towards the design of static anti-windup gains arising from suitable performance/robustness trade-offs. More specifically, after analyzing in-depth the formal properties and the algorithmic solutions for the novel randomized approach, we apply it to the robust design of anti-windup compensators within different problem settings; namely, we address the minimization of the nonlinear \Ltwo\ gain, the minimization of the area spanned by the nonlinear \Ltwo\ gain curve, the minimization of the reachable set and the maximization of the domain of attraction for closed-loop saturated systems. For each of the above problems, we provide several discussions and a suitable simulation example (in Section~\ref{sec:example}). 

The paper is ended by some concluding remarks.

\subsection*{Notation}
In the remainder of the paper, the following notation and definitions are adopted:
\begin{itemize}
\item the $\mathcal{L}_2$ norm of a scalar valued signal $x(t)$, defined for $t \geq 0$, is
\[\Vert x_2\Vert \doteq \left(\int_{0}^{\infty}x^2(t) \ dt\right)^{1/2} ;\] 
\item $\mathrm{e}$ denotes the Euler number;
\item given a square matrix $Z$, $\He(Z) \doteq Z+Z^T$;
\item given a matrix $X$, $X_{[k]}$ denotes the $k^{th}$ row of $X$.
\end{itemize}

%%%%%%%%%%%%%%%%%%%%%%%%%%%%%%%%%%%%%%%%%%%%%%%%%%%%%%%%%%%%%%%%%%%%%%%%%%%%%%%%
%%%%%%%%%%%%%%%%%%%%%%%%%%%%%%%%%%%%%%%%%%%%%%%%%%%%%%%%%%%%%%%%%%%%%%%%%%%%%%%%
\section{Scenario with certificates}\label{sec:SwC_big}

In this section, we briefly recall the scenario approach in dealing with convex optimization problems in the presence of uncertainty, and subsequently introduce a novel framework that we name {\textit{scenario with certificates} (SwC)}.

\subsection{The scenario approach}
The so-called scenario approach \cite{CalCam:06tac} has been developed  to deal with robust convex optimization problems of the form 
\begin{align}
\tRO  =  \arg &\displaystyle \min_{\theta\in\Theta} c^T \theta     \tag{RO}\label{eq:robust_opt} \\
                  &    \text{s.t. }   f(\theta,q) \leq 0,    \    \forall q  \in \Q,\nonumber
\end{align}
where, for given $q$ within the uncertainty set $\Q$, $f(\theta,q)$ are convex functions of the optimization variable $\theta\in\Theta$, the domain $\Theta$ is a convex and compact set in $\real^{n_{\theta}}$ and the uncertainty set $\Q$ is not necessarily compact.
Furthermore, we assume that $f(\theta,q)$ is a continuous (possibly nonlinear) function of $q$ for any given $\theta$.

Following the probabilistic approach discussed for instance in \cite{TeCaDa:13,CaDaTe:11} a probabilistic description of the uncertainty is considered over $\Q$. That is, we formally assume that $q$ is a random variable with given probability distribution with support $\Q$. Such a probability distribution may describe the likelihood of each occurrence of the uncertainty or a user-defined weight for all possible uncertain situations. Then, $N$ independent identically distributed (iid) samples $q^{(1)}, \ldots , q^{(N)}$ are extracted according to the probability distribution of the uncertainty over $\Q$.

These samples are used to construct the following scenario optimization (SO) problem,
based on $N$ instances (scenarios) of the uncertain constraints 
\begin{align}
         \tSO    =   \arg &\displaystyle\min_{\theta\in\Theta} c^T \theta    \tag{SO}\label{eq:scenario_opt} \\
                    & \text{s.t. }     f(\theta,q ^{(i)}) \leq 0, \  i=1,\ldots,N. \nonumber
\end{align}
Problem \eqref{eq:scenario_opt} can be seen as a probabilistic relaxation of  problem \eqref{eq:robust_opt}, since it deals only with a subset of the constraints considered in \eqref{eq:robust_opt}, according to the probability distribution of the uncertainty. However, under rather mild assumptions on problem \eqref{eq:robust_opt}, by suitably choosing~$N$, this approximation may in practice become negligible in some probabilistic sense. Specifically, $N$ can be selected depending on the level of ``risk'' of constraint violation that the
user is willing to accept. 
To this end, the \textit{violation probability} of the design $\theta$ is defined as
\begin{equation}
\label{eq:viol}
\V(\theta)\doteq\Pr\left\{q\in\Q \,:\,f(\theta,q)>0\right\}
\end{equation}
{where $\Pr$ denotes the probability with respect to the distribution of the random variable $q$}.
Similarly,  the \textit{reliability} of the design $\theta$ is given by
\[
\R(\theta)\doteq1-\V(\theta).
\]
Then the following result has been proven in \cite{CamGar:08}.
\begin{proposition}
\label{prop:scenario}
\cite{CamGar:08} Assume that, for any multisample extraction, problem (\ref{eq:scenario_opt}) is feasible and attains a unique optimal solution. Then,  given an accuracy level $\epsilon\in(0,1)$, the solution $\tSO$ of problem \eqref{eq:scenario_opt} satisfies
\begin{equation}\label{eq:binom}
\Pr\left\{\V(\tSO)>\epsilon\right\} \le \B(N,\epsilon,n_{\theta}),
\end{equation}
where
\begin{equation}\label{eq:binom2}
\B(N,\epsilon,n_{\theta}) \doteq \sum_{k=0}^{n_{\theta}-1}\binom{N}{k}\epsilon^{k}(1-\epsilon)^{(N-k)}.
\end{equation}
\end{proposition}

\smallskip

We note that non-uniqueness of the optimal solution can be circumvented by imposing additional ``tie-break'' rules in the problem, see, \textit{e.g.}, Appendix A of \cite{CalCam:06tac}.
Also, in \cite{Calafiore:10siopt} it is shown that the feasibility assumption can be removed at the expense of substituting $n_{\theta}-1$ with $n_{\theta}$ in $\B(N,\epsilon,n_{\theta})$.

From Equation (\ref{eq:binom}), explicit bounds on the number of samples necessary to guarantee the  ``goodness'' of the solution have been derived. The bound provided in \cite{ATLR:15} shows that, if, for given $\epsilon,\delta\in(0,1)$, the sample complexity $N$ is chosen  to satisfy the  bound
\begin{equation}\label{eq:N}
N\geq \frac{\mathrm{e}}{\epsilon(\mathrm{e}-1)}\left(\ln\frac{1}{\delta}+n_{\theta}-1\right)m,
\end{equation} 
then the solution $\tSO$ of problem \eqref{eq:scenario_opt} satisfies $\V(\tSO)\le\epsilon$ with probability $1-\delta$. This bound improves  by a constant factor upon previous bounds, see e.g.~\cite{Calafiore:10siopt}, and it  shows that problem (SO) exhibits linear dependence in $1/\epsilon$ and $n_{\theta}$, and logarithmic dependence on $1/\delta$. Note however that, from a practical viewpoint, it is always preferable to numerically solve the one dimensional problem of finding the smallest {integer} $N$ such that $\B(N,\epsilon,n_{\theta})\le\delta$.

\smallskip

\subsection{Scenario with certificates}
The classical scenario approach previously discussed deals with uncertain optimization problems where  all  variables $\theta$ are to be designed. 
On the other hand,  in the design with certificates approach we distinguish between \textit{design variables}~$\theta$ and \textit{certificates} $\xi$. In particular, we consider now a function $f(\theta,\xi,q)$, which {is assumed to be \textit{jointly convex}} in $\theta\in\Theta$ and $\xi\in\Xi \subseteq \mathbb{R}^{n_\xi}$ for given $q\in\Q$ (where $\Theta$ and $\Xi$ are supposed to be non-empty), and construct the following robust optimization problem with certificates 
% ROC
\begin{align}
\tROC  =   \arg &\displaystyle\min_{\theta} c^T \theta \tag{RwC}\label{eq:certificates_opt}\\
                  &  \text{s.t. }      \theta \in \S(q),    \    \forall q  \in \Q, \nonumber
\end{align}
where the set $\S(q)$ is defined as
\begin{equation}
\S(q)\doteq
\left\{\theta\in\Theta | \ \exists \xi\in\Xi
\text{ satisfying } f(\theta,\xi,q)\le 0
\right\}.
\label{eq:S_of_q}
\end{equation}
{The key observation that is at the basis of the approach developed in this section is that  the set $\S(q)$ is convex in~$\theta$ for any given~$q$, as formally shown in Theorem 1 below.}

\begin{remark}[Common vs. parameter-dependent certificates]
As discussed in the Introduction,  problem \eqref{eq:certificates_opt} corresponds to searching for  so-called \textit{parameter-dependent} certificates, in the sense that a different certificate is allowed for every instance of the uncertainty $q$, that is $\xi=\xi(q)$. This is very different from the approach frequently adopted when dealing with uncertain systems, based on the design of \textit{common} certificates. 
This would result in a robust problem of the form
\begin{align}
\{\theta\ped{CO},\xi\ped{CO}\}= \arg &\displaystyle \min_{\theta\in\Theta,\xi\in\Xi} c^T \theta     \tag{CO}\label{eq:common_opt} \\
                  &    \text{s.t. }   f(\theta,\xi,q) \leq 0,    \    \forall q  \in \Q,\nonumber
\end{align}
where the common certificate $\xi\ped{CO}$ should be the same for all possible values of $q$. Clearly, if the spread of the uncertainty is large, it is unreasonable to expect the same certificate $\xi\ped{CO}$ to hold for all $q\in\Q$.
For instance, in the classical case when the certificates correspond to Lyapunov functions for proving stability, the difference between the two approaches lies on the difference between common Lyapunov functions and parameter-dependent ones. In particular, for this problem, different solutions have been proposed in the robust control literature, which are based on explicit parameterizations (e.g. linear or bilinear) of the function $\xi(q)$, see for instance \cite{Barmish:94}. One of the main novelties of the probabilistic approach discussed in this paper is the fact that no explicit parameterization is necessary.
\end{remark} 

In \cite{Oishi:07a}, an approach to handle parameter-dependent  linear matrix inequalities (LMIs) has been introduced, and a solution for feasibility problems, based on uncertainty randomization and on an iterative ellipsoidal algorithm, has been derived. The approach considers different certificates for each sampled value of the random uncertainty. In the same paper, the conservatism reduction is illustrated by means of a numerical example showing that traditional robustness methods based on common Lyapunov functions fail.

In the current work, we follow along this line of research, and propose to approximate problem \eqref{eq:certificates_opt} introducing the following \textit{scenario  with certificates} problem, based again on a multisample extraction
%SwC
\begin{align}
         \tSOC    =   \arg&\displaystyle \min_{\theta,\xi_{1},\ldots,\xi_{N}} c^T \theta  \tag{SwC}\label{eq:scenario_cert}\\
                    &  \text{s.t. }  f(\theta,\xi_{i},q^{(i)}) \leq 0, \  i=1,\ldots,N.\nonumber
\end{align}
Note that, contrary to problem \eqref{eq:scenario_opt}, in this case a new certificate variable $\xi_{i}$ is created for every sample $q^{(i)}$, $i=1,\ldots,N$, that is $\xi_i=\xi_i(q^{(i)})$. To analyze the properties of the solution $\tSOC$, we note that, in the case of SwC, the reliability and violation probabilities of design $\theta$ are given by
\begin{eqnarray*}
\R(\theta)&=&
\Pr\Bigl\{q\in\Q | \exists \xi\in\Xi \text{ satisfying } f(\theta,\xi,q)\le 0 \Bigr\},\\
\V(\theta)&=&
\Pr\Bigl\{\exists q\in\Q  | \nexists \xi\in\Xi \text{ satisfying } f(\theta,\xi,q)\le 0 \Bigr\}.
\end{eqnarray*}

We now state the main result regarding the scenario optimization with certificates.
\begin{theorem}
\label{them:SwC}
Assume that, for any multisample extraction, problem \eqref{eq:scenario_cert} is feasible and attains a unique optimal solution.
Then, given an accuracy level $\epsilon\in(0,1)$, the solution $\tSOC$ of problem \eqref{eq:scenario_cert} satisfies
\begin{equation}\label{eq:binom_reprise}
\Pr\left\{\V(\tSOC)>\epsilon\right\}\le\B(N,\epsilon,n_{\theta}).
\end{equation}
\end{theorem}
\smallskip

\proof
{We first prove convexity of the set $\S(q)$. To see this, consider $\theta_{1},\theta_{2}\in\S(q)$. 
Then, there exist $\xi_{1},\xi_{2}$ such that 
\[f(\theta_{1},\xi_{1},q)\le 0 \text{ and } 
f(\theta_{2},\xi_{2},q)\le 0.
\] Consider now  $\theta_{\lambda}\doteq \lambda \theta_{1}+(1-\lambda)\theta_{2}$, with $\lambda\in[0,\,1]$, and let $\xi_{\lambda}=\lambda \xi_{1}+(1-\lambda)\xi_{2}$. From convexity of $f$ with respect to both $\theta$ and $\xi$ it immediately follows that
\[
f(\theta_{\lambda},\xi_{\lambda},q)\le \lambda f(\theta_{1},\xi_{1},q) + (1-\lambda)f(\theta_{2},\xi_{2},q)\le 0,
\]
hence $\theta_{\lambda}\in\S(q)$, which proves convexity.}

Now, observe that the condition $\theta\in\S(q)$ is equivalent to requiring 
\[
f_{\xi}(\theta,q)\doteq\inf_{\xi\in\real^{n_{\xi}}}
f(\theta,\xi,q)\le 0,
\]
so that problem \eqref{eq:certificates_opt} is equivalent to
\begin{eqnarray}\label{eq:certificates_opt2}
 && \min_{\theta} c^T \theta \\
 & & \mbox{s.t. }   f_{\xi}(\theta,q)\le 0 \quad    \forall q  \in \Q.\nonumber
\end{eqnarray}
{Note that, from the convexity of $\S(q)$, it follows that the function $f_{\xi}(\theta,q)$ is convex in $\theta$  for given $q$;}
see also \cite[p.\ 113]{BoyVan:04}. Hence, problem \eqref{eq:certificates_opt2} is a robust convex optimization problem. Then, we construct its scenario counterpart
\begin{eqnarray}
&& \min_{\theta} c^T \theta,  \label{eq:scen2}\\
& & \mbox{s.t. } \min_{\xi_{i}\in\real^{n_{\xi}}} f(\theta,\xi_{i},q^{(i)}) \le 0, \, i=1,\ldots,N, \nonumber
\end{eqnarray}
where the subscript $i$ for the variables $\xi_{i}$ highlights that the different minimization problems are independent. Finally, we note that \eqref{eq:scen2} immediately rewrites as problem \eqref{eq:scenario_cert}.
\qed
\vskip .3cm

We remark that problem (SwC) has $N$ separate constraints, one for each $q^{(i)}$, and each constraint involves a different certificate. However, notice that the dimension $n_\xi$ of the certificates~$\xi$ does not enter in the right-hand side of the probability bound \eqref{eq:binom_reprise} in Theorem~\ref{them:SwC}. Hence, the sample complexity of problem (SwC)
is smaller than that of the scenario counterpart of the problem with common certificates (CO), in which both $\theta$ and $\xi$ play the role of design variables. On the other hand, the complexity of solving problem (SwC) is  higher, since the number of optimization variables significantly increases, because a different variable $\xi_i$ is introduced for every sample $q^{(i)}$. This increase in complexity is not surprising, being problem (RwC) much more difficult than problem (CO). 
In particular, we remark that, in the case when the constraints are linear matrix inequalities, then the scenario problem can be reformulated as a semidefinite program by combining the~$N$ LMIs into a single LMI with block-diagonal structure. It is known, see \cite{BenNem:01}, that the computational cost of this problem with respect to the number of diagonal blocks $N$ is of the order of $N^{3/2}$. The sequential method discussed in the next section aims at improving the computational efficiency by reducing the number of scenarios. 

\subsection{Sequential randomized algorithm for SwC}\label{sec:SwC}

{Motivated by the computational burden of the SwC solution, 
in this section, we  present a sequential randomized algorithm that alleviates the load 
by solving a series of reduced-size problems.}
The algorithm is a minor modification of \cite[Algorithm 1]{CDTV:15}, which was introduced for the standard scenario approach, and it is based on separate design and validation steps. The design step requires the solution of the reduced-size SwC problem. In the validation step, contrary to \cite{CDTV:15} where only functional evaluations are required, the feasibility problems \eqref{eq:SwC alg1 design} and \eqref{eq:SwC alg1 valid} need to be solved. However, it should be pointed out that the latter problems are of small size, and can be solved independently, and hence parallelized.
The sequential procedure  is presented in Algorithm 1, and its theoretical properties are  stated in the subsequent lemma. 
Its proof follows the same lines of that in \cite[Theorem 1 and Algorithm 1]{CDTV:15}, and is omitted for brevity. It should be stressed, however, that in \cite{CDTV:15} the sequential approach was not applied to SwC, but to standard scenario optimization.

%%%%%%%%%%%%%%%%%%%%%%%%%%%%%%%%%%%%%%%%%%%%%%%%%%%%%%%%%%%%%%%%%%%%%%%%%%%%%%%%%%%%%%%%%%%%%
\noindent \line(1,0){250}
\begin{center}
\textbf{Sequential Algorithm for SwC}
\end{center}
\vspace{-1.5mm}
\line(1,0){250}
\begin{enumerate} 
  \item \textsc{Initialization}\newline 
	set the iteration counter $k=0$. Choose the desired probabilistic levels $\epsilon$, $\delta$ and the desired number of iterations $k_t>1$
  \item \textsc{Update}\label{item:update}\newline
  set $k=k+1$ and $N_k\ge N\frac{k}{k_t}$ where $N$ is the smallest integer s.t. 
  $\B(N,\epsilon,n_{\theta})\leq\delta/2$
  \item \textsc{Design}
  \begin{itemize}
    \item draw $N_k$ iid (design) samples
    $\{q_d^{(1)},\ldots, q_d^{(N_k)}\}$
    \item solve the following \textit{reduced-size SwC problem}
\begin{eqnarray}\label{eq:SwC alg1 design}
        & \hat{\theta}_{N_k}    =&\!\!\!\!   \arg\displaystyle \min_{\theta,\xi_{1},\ldots,\xi_{N_k}} c^T \theta,       \\
          &          &\!\!\!\!\mbox{s.t. }     f(\theta,\xi_{i},q_{d}^{(i)}) \leq 0, \  i=1,\ldots, N_k.\nonumber
\end{eqnarray}
    \item {if} the last iteration is reached $(k=k_t)$, \\{return} $\tSSOC=\hat{\theta}_{N_k}$
  \end{itemize}
  \item \textsc{Validation}
  \begin{itemize}
    \item set $M_{k}$ according to \eqref{eq:sample bound Mk}
    \item draw iid (validation) samples $ \{q_v^{(1)},\ldots, q_v^{(M_k)}\}$
    \item {for} $j=1$ {to} $M_{k}$
  \begin{itemize}
    \item {if} the validation problem
\begin{eqnarray}\label{eq:SwC alg1 valid}
         &&   \text{find } \xi_{j} \mbox{ such that}     \nonumber \\
                    &   &  f(\hat{\theta}_{N_k},\xi_{j},q_{v}^{(j)}) \leq 0    
\end{eqnarray}
        is unfeasible  {goto} step (\ref{item:update}).
\end{itemize}
\item 
{return} $\tSSOC=\hat{\theta}_{N_k}$.
  \end{itemize}
\end{enumerate}
\vspace{-1.5mm}
\line(1,0){250}
%%%%%%%%%%%%%%%%%%%%%%%%%%%%%%%%%%%%%%%%%%%%%%%%%%%%%%%%%%%%%%%%%%%%%%%%%%%%%%%%%%%%%%%%%%%%%

\begin{lemma}\label{theo:property of algorithm 1}
Assume that, for any multisample extraction, problem \eqref{eq:SwC alg1 design} is feasible and attains a unique optimal solution. Then,  given  accuracy level $\epsilon\in(0,1)$ and confidence level $\delta\in(0,1)$, let
    \begin{equation}\label{eq:sample bound Mk}
    M_k\ge\frac{\alpha\ln k+\ln \left(\mathcal{H}_{k_t-1}(\alpha)\right)+\ln\frac{2}{\delta}}{\ln\left(\frac{1}{1-\epsilon}\right)}
    \end{equation}
where $\mathcal{H}_{k_t-1}(\alpha)=\sum_{j=1}^{k_t-1}j^{-\alpha}$, with $\alpha>0$, is a finite hyperharmonic series.
Then, the probability that at iteration $k$  Algorithm 1 returns a solution $\tSSOC$
with violation greater than $\epsilon$ is at most $\delta$, \textit{i.e.},
\begin{equation}\label{eq:thr2}
\Pr\left\{\V(\tSSOC)>\epsilon\right\}\le\delta.
\end{equation}
\end{lemma}

\begin{remark}The dimension of the system that can be handled by the algorithm depends not only on $n_{\theta}$, but also on the desired probabilistic accuracy and confidence. The paper \cite{CDTV:15} considers a real-world example of a hard disk drive consisting of $153$ design parameters and $9$ uncertain parameters. It is shown that the sequential approach provides results even for very tiny values of accuracy and confidence, contrary to the one-shot solution, i.e.\  the one considering all the $N$ constraints at once.\end{remark}

In the second part of this paper, we introduce the problem of robust \Ltwo\ gain minimization for linear anti-windup systems. The SwC approach appears to be well suited for such a design problem, for several reasons: i) the nominal design can be formulated in terms of linear matrix inequalities, ii) the uncertainty set can in principle be of any size and shape, and iii) the optimization variables can be easily divided in design variables for the anti-windup augmentation and certificates for stability and performance guarantees, iv) the number of uncertain parameters can in principle be arbitrarily large and any functional dependence is allowed.

%%%%%%%%%%%%%%%%%%%%%%%%%%%%%%%%%%%%%%%%%%%%%%%%%%%%%%%%%%%%%%%%%%%%%%%%%%%%%%%%
\section{Anti-windup compensator design}\label{ref:aw}
Consider the linear uncertain continuous-time plant {with $n_u$ inputs subject to saturation}
\begin{eqnarray}
\label{eq:P}
\dot{x}_p         & =  &  A_p(q) x_p       +  B_{p,u}(q)\sigma   +  B_{p,w}(q)w \nonumber \\
    y             & =  &  C_{p,y}(q) x_p   +  D_{p,yu}(q)\sigma  +  D_{p,yw}(q)w \\
    z             & =  &  C_{p,z}(q) x_p   +  D_{p,zu}(q)\sigma  +  D_{p,zw}(q)w, \nonumber
\end{eqnarray}
where $x_p$ is the plant state, {$\sigma\in\mathbb{R}^{n_u}$} is the control input, $w$ is an external input (possibly comprising references and disturbances), $z$ is the performance output, $y$ is the measured output and $q$ denotes random uncertainty within the set $\Q$. We denote by $\bar q\in\Q$ the nominal value of the uncertain parameters.

As customary with linear anti-windup design \cite{ZackAWbook11}, we assume that a linear controller has been designed, based on the nominal system, in order to induce suitable nominal closed-loop properties when interconnected to plant (\ref{eq:P})
\begin{eqnarray}
\label{eq:C}
\begin{array}{rcl}
\dot{x}_c   & =  &  A_c x_c       +  B_{c,y}y   +  B_{c,w}w + v_1  \\
    u       & =  &  C_{c} x_c +  D_{c,y}y +  D_{c,w}w + v_2,
\end{array}
\end{eqnarray}
where $x_c$ is the controller state, $w$ typically comprises
references (but may also contain disturbances), $u$ is the controller
output and $v= [v_1^T \ v_2^T]^T$ is an extra input available for
anti-windup action. The controller (\ref{eq:C}) is typically designed in
such a way that the so-called {\em unconstrained closed-loop system}
given by (\ref{eq:P}), (\ref{eq:C}), $\sigma=u$, $v=0$ is nominally
asymptotically stable and satisfies some nominal or robust performance requirements. 

Consider now the (physically more reasonable) {\em saturated interconnection} $\sigma=\sat(u)$, where the $k^{th}$ entry of $\sigma$ is $\sat_k (u_k) = \max(\min(\bar u_k,u_k),-\bar u_k)$, denoting the $k^{th}$ input by $u_k$. When the input saturates, the closed loop system composed by the feedback loop between (\ref{eq:P}) and (\ref{eq:C}) is no longer linear and may exhibit undesirable behavior, usually called {\em controller windup}. Then, one may wish to use the free input $v$ to design a suitable {\em static anti-windup compensator} of the form 
\begin{eqnarray}
\label{eq:static_aw}
v = [v_1^T \ v_2^T]^T = D_{aw} (u - \sat(u)).
\end{eqnarray}
This signal can be injected into the right hand side of the controller dynamics (\ref{eq:C}) to recover stability and performance of the unconstrained closed-loop system.

When lumping together the plant-controller-anti-windup components (\ref{eq:P}), (\ref{eq:C}), (\ref{eq:static_aw}), $\sigma=\sat(u)$, one obtains the so-called {\em anti-windup closed-loop system}, a nonlinear control system which can be compactly written using the state $x = [x_p^T \ x_c^T]^T$ as in (\ref{eq:CL}) (at the top of the next page),
\begin{figure*}
\begin{eqnarray}
\label{eq:CL}
\begin{array}{rcl}
\dot{x}        & =  &  A_{cl}(q) x   +  \left(B_{cl,q}(q)+B_{cl,v}(q)D_{aw}\right)\dz(u)    +  B_{cl,w}(q)w \\
    z          & =  &  C_{cl,z}(q) x +  \left(D_{cl,zq}(q)+D_{cl,zv}(q)D_{aw}\right)\dz(u)  +  D_{cl,zw}(q)w \\
    u          & =  &  C_{cl,u}(q) x +  \left(D_{cl,uq}(q)+D_{cl,uv}(q)D_{aw}\right)\dz(u)  +  D_{cl,uw}(q)w
\end{array}
\end{eqnarray}
\begin{center}
\line(1,0){500}
\end{center}
\begin{subequations}
\label{eq:analysisALL}
\begin{align}
\label{eq:analysis1}
    &  Q =Q^T>  0, \; U  > 0 \mbox{ diagonal},     \\  
\label{eq:an_bigLMI}
       &   \He\begin{bmatrix}
              A_{cl}(q)Q        &      \left(B_{cl,q}(q)+B_{cl,v}(q)D_{aw}\right)U+  Y^T            &   B_{cl,w}(q) & 0          \\
	              C_{cl,u}(q)Q    &     \left(D_{cl,uq}(q)+D_{cl,uv}(q)D_{aw}\right)U-U         &  D_{cl,uw}(q) & 0         \\
	                    0            &      0               &   -I/2 & 0\\
	              C_{cl,z}(q)Q    &      \left(D_{cl,zq}(q)+D_{cl,zv}(q)D_{aw}\right)U           &    D_{cl,zw}(q)   & -\gamma^2 I/2
	\end{bmatrix} < 0,\\	
&        \label{eq:analysis2}      \begin{bmatrix}
              Q       &      Y_{[k]}^T          \\
	    Y_{[k]}            &     \bar{u}_k^2/s^2
	\end{bmatrix} \geq 0, \quad k=1,\ldots, n_u,
\end{align}
\end{subequations}
\begin{center}
\line(1,0){500}
\end{center}
\begin{equation}
\label{eq:synthesis}
\begin{array}{rl}
& \hspa \He\!\! \begin{bmatrix}
              A_{cl}(q)Q        &      B_{cl,q}(q)U +B_{cl,v}(q)X +  Y^T      &   B_{cl,w}(q) 		& 0          \\
						C_{cl,u}(q)Q\!\!    &     D_{cl,uq}(q) U +D_{cl,uv}(q)X -U        &  D_{cl,uw}(q) 		& 0         \\
								0            &      								0	               &   -I/2 				& 0\\
				C_{cl,z}(q)Q \! \!      &      D_{cl,zq}(q)U +D_{cl,zv}(q)X           &    D_{cl,zw}(q)   &\!\!\! -\frac{\gamma^2}{2}I
	\end{bmatrix}\! < \! 0
\end{array}
\end{equation}
\begin{center}
\line(1,0){500}
\end{center}
\end{figure*}
where $\dz$ denotes the deadzone function, \textit{i.e.}, $\dz(u) = u- \sat(u)$, and all the matrices are uniquely determined by the data in (\ref{eq:P}), (\ref{eq:C}), (\ref{eq:static_aw}) (see, \textit{e.g.}, the full authority anti-windup section in \cite{ZackAWbook11} for explicit expressions of these matrices).

The compact form in (\ref{eq:CL}) may be used to represent both the saturated closed loop before anti-windup compensation,  by selecting $D_{aw}=0$,  or the closed loop with anti-windup compensation,  by performing some nonzero selection of~$D_{aw}$. 

\subsection{$\mathcal{L}_2$ gain minimization}

First, we analyze system (\ref{eq:CL}) for the \textit{nominal case}, that is when no uncertainty is present and $\Q$ is a singleton
coinciding with the nominal value $\bar q$ of the parameters. 

In this nominal case, the results in \cite{TarbouriechTAC05,HuAuto08,HuTAC06,TarBook11} and references therein generalize the well-known sector conditions originating from absolute stability theory, into a so-called generalized sector condition, stating that given any matrix $H$, it holds that $\dz(u)^T U^{-1} (u-\dz(u)+Hx) \geq 0$
for all $x$ satisfying $\dz(Hx)=0$.
% along solutions to (\ref{eq:CL}). 
This condition is a powerful tool because it enables us to provide a non-global homogeneous characterization of the stability and performance properties of the
nonlinear closed loop (\ref{eq:CL}) by way of an extension of absolute stability theory.
 In particular, 
in (\ref{eq:analysisALL})
the generalized sector condition provides guarantees on the derivative of a quadratic Lyapunov function
$x^TQ^{-1}x$ in a suitable (ellipsoidal) sublevel set ${\mathcal E}((s^2Q)^{-1})$
(see (\ref{eq:sublevelset}) below)
contained in the region where $\dz(Hx)=0$ (this is guaranteed by \eqref{eq:analysis2}). Here, parameters $Q$, $U$
and $Y=U^{-1}H$ can be optimized by way of a convex semi-definite program.
More formally, we recall the following stability and performance analysis result from \cite[Theorem 2]{HuTAC06}.

\begin{proposition}[Regional stability/performance analysis]
\label{prop:AWanalysis}
Given a scalar $s>0$, consider the nominal system, that is let $\Q\equiv\{\bar q\}$.
Assume that the semidefinite programming (SDP) problem \eqref{eq:analysisALL} in the variables $\gamma^2$, $Q$, $Y$ and $U$ is feasible. Then:
\begin{enumerate}[(a)]
	\item the nonlinear algebraic loop in (\ref{eq:CL}) is well posed,
	\item the origin is locally exponentially stable for (\ref{eq:CL}) with
basin of attraction containing the set 
\begin{equation}
\label{eq:sublevelset}
{\mathcal E}((s^2Q)^{-1}) = \{x:\; x^T Q^{-1} x \leq s^2\},
\end{equation}
 \item for each $w$ satisfying $\|w\|_2\leq s$, the zero initial state solution to (\ref{eq:CL}) satisfies $\|z\|_2 \leq \hat{\gamma} \|w\|_2$, where the \Ltwo\ gain of the system is given by
	\begin{eqnarray}\label{eq:L2_computation_analysis}
	\hat{\gamma}^2(s) &=& \displaystyle\min_{\left\{\gamma^2,Q,Y,U\right\}}\gamma^2 \\
	 &&\mbox{\rm s.t. (\ref{eq:analysisALL}).} \nonumber
	\end{eqnarray}
\end{enumerate}
\end{proposition}
\smallskip
As suggested in \cite{HuTAC06}, one may use the result of
Proposition~\ref{prop:AWanalysis} to compute an estimate of the nominal nonlinear \Ltwo\ gain curve (see \cite{Megretski96}), namely a function $s\mapsto \hat \gamma(s)$ such that for each $s$ in the feasibility set of (\ref{eq:analysisALL}) and for each $w$ satisfying $\|w\|_2 \leq s$, the
zero initial state solution to (\ref{eq:CL}) satisfies 
$$
\|z\|_2 \leq \hat \gamma(s) \|w\|_2.
$$
To do so, it is possible to sample the nonlinear gain curve $s \mapsto \hat \gamma(s)$ by selecting suitable positive values $s_1 < \cdots< s_n$ and, for each $k=1,\ldots, n$, solving  (\ref{eq:L2_computation_analysis}), after replacing $s= s_k$.
Then, the \Ltwo\ gain curve estimate can be constructed by interpolating the points $(s_k,\hat{\gamma}_k(s_k))$, $k=1,\ldots, n$.

Following the derivations in \cite{TarbouriechTAC05} (which generalize the global results of \cite{MulderAuto01}), one may notice that the product $D_{aw} U$ appears in a linear way in equation (\ref{eq:an_bigLMI}) and, for a fixed value of $s$, the synthesis of a static anti-windup gain minimizing the nonlinear \Ltwo\ gain can be written as a convex optimization problem, as stated next.

\begin{proposition}[Regional stability/performance synthesis]
\label{prop:AWsynthesis}
Given the plant-controller pair (\ref{eq:P}), (\ref{eq:C}), and a scalar $s>0$,
consider the nominal system, that is $\Q\equiv\{\bar q\}$ is a singleton. Assume that the SDP problem
\begin{eqnarray}\label{eq:L2_computation_synthesis}
	\hat{\gamma}^2(s) &=& \displaystyle\min_{\left\{\gamma^2,Q,Y,U,X\right\}}\gamma^2 \\
	 &&\mbox{\rm s.t. }(\ref{eq:analysis1}), (\ref{eq:analysis2}), (\ref{eq:synthesis}) \nonumber
	\end{eqnarray}
is feasible.
Then, selecting the static anti-windup gain as
\begin{equation}
\label{eq:staticAWsel}
D_{aw}=XU^{-1},
\end{equation}
the anti-windup closed-loop system (\ref{eq:P}), (\ref{eq:C}), (\ref{eq:static_aw}), $\sigma=\sat(u)$ or its equivalent representation in (\ref{eq:CL}) satisfies properties (a)-(c) of Proposition~\ref{prop:AWanalysis}.
\end{proposition}
\smallskip

\begin{remark}
The static linear anti-windup architecture (\ref{eq:static_aw}) adopted in 
Proposition~\ref{prop:AWsynthesis} and in the rest of this paper
 is only one among many possible choices (see, e.g., \cite{ZackAWbook11}). 
In particular, when using direct linear anti-windup designs, an alternative appealing approach is given by the design of a plant-order linear filter (namely, of the same order of the plant) generalizing the static selection in (\ref{eq:static_aw}). 
Such a dynamic generalization
of (\ref{eq:static_aw}) was shown in \cite{Grimm03TAC} to be important to guarantee global exponential stability in the presence of saturation. However, this fact was later de-emphasized once the above mentioned generalized sector condition was introduced (see \cite{HuAuto08}, which provides the non-global extension of the results in \cite{Grimm03TAC}). More specifically, non-global guarantees of stability in the presence of saturations/deadzones is a fundamental tool to establish exponential stability properties of the origin for non-asymptotically stable plants that are stabilized through a saturated control input.
\end{remark}

\begin{remark}
The separation between the Lyapunov certificate $Q,Y$ and the optimization variables $X,U$ in (\ref{eq:synthesis}) is only possible when adopting the static architecture in (\ref{eq:static_aw}), which makes the robust extensions provided below reasonably simple. Extensions to the dynamic plant-oder anti-windup case is possible only if one adopts certain conservative convex relaxations of the nonconvex robust conditions, along similar directions to those well surveyed, for example, in \cite{ebihara2015s}.
\end{remark}
%%%%%%%%%%%%%%%%%%%%%%%%%%%%%%%%%%%%%%%%%%%%%%%%%%%%%%%%%% 

Consider now the \textit{uncertain case}, when the system matrices in \eqref{eq:CL} defining the dynamics of $x$ and $z$ are continuous (possibly nonlinear) functions of the uncertainty $q \in \Q$, which is considered to be time invariant. Then, the interest is in finding robust solutions to the analysis and design problems discussed before. For instance, in the analysis case, one could search for common certificates $Q,Y,U$ in \eqref{eq:L2_computation_analysis} such that $\gamma^{2}$ is minimized over (\ref{eq:analysisALL}) for all $q\in\Q$. This approach is pursued in \cite{FormentinCDC13}, where scenario results are used to find probabilistic guaranteed estimates. 
Note that the use of a common Lyapunov function is well justified when the uncertainty is, for instance, time-varying.
However, as discussed in Section~\ref{sec:SwC}, an approach based on common certificates is in general very conservative in the case of time-invariant uncertainty, and one would be more interested in finding \textit{parameter-dependent} certificates. To do this, we would need to solve the
following robust optimization problem with certificates
\begin{align}
\hat{\gamma}^2(s) =& \displaystyle\min\gamma^2 \label{eq:RwC-AWA} \\
&\text{s.t. } \gamma^{2}\in
\left\{\gamma^{2} | \ \exists 
\{Q,Y,U\}
\text{ satisfying }\eqref{eq:analysisALL}
\right\} \,\forall q\in\Q.\nonumber
\end{align}
A similar rationale can be applied to robustify the anti-windup synthesis problem of Proposition~\ref{prop:AWsynthesis}. As a matter of fact, when the system matrices are uncertain, one meets similar obstructions to those highlighted as far as analysis was concerned. Again, instead of looking for common Lyapunov certificates as in \cite{FormentinCDC13}, we write the following RwC problem
\begin{align}
\hat{\gamma}^2(s) =& \displaystyle\min\gamma^2 &\label{eq:RwC-AWS} \\
&\text{s.t. } \{\gamma^{2},U,X\}\in \bigl\{ \{\gamma^{2},U,X\}\  | \ \exists 
\{Q,Y\}
\text{ satisfying } \nonumber \\ & (\ref{eq:analysis1}), (\ref{eq:analysis2}), (\ref{eq:synthesis}) \bigr\} \quad\forall q\in\Q.\nonumber
\end{align}

Note that both problems \eqref{eq:RwC-AWA} and \eqref{eq:RwC-AWS} are difficult nonconvex semi-infinite optimization problems, due to the fact that one has to determine the certificates as functions of the uncertain parameter $q$. A classical approach in this case is to assume a specific dependence (generally affine) of the certificates on the uncertainty.
Instead, in this paper we adopt  a probabilistic approach, assuming that $q$ is a random variable with given probability distribution over $\Q$, and apply the SwC approach discussed in Section~\ref{sec:SwC}. This allows us to find an implicit dependence on $q$ of the certificates. This is in the spirit of the original idea proposed in \cite{Oishi:07a}. 
The following two theorems, whose proofs come straightforwardly from Propositions~\ref{prop:scenario} and~\ref{prop:AWanalysis}, exploit the SwC approach to address the robust nonlinear \Ltwo\ gain estimation  and synthesis for saturated systems. 

\begin{figure*}
\begin{equation}
\label{eq:synthesis-SwC}
\begin{array}{rl}
 \hspa \hat{\gamma}^2(s) =& \min\nolimits\limits_{\left\{\gamma^2,Q_{1},\ldots,Q_{N},Y_{1},\ldots,Y_{N},U,X\right\}}\gamma^2 \\
& \hspa \mbox{s.t. } 
  Q_{i} =Q_{i}^T>  0, \; U  > 0 \mbox{ diagonal},      \\  
& \quad \hspa \He\!\! \begin{bmatrix}
              A_{cl}(q^{(i)})Q_i        &      B_{cl,q}(q^{(i)})U +B_{cl,v}(q^{(i)})X +  Y_i^T      &   B_{cl,w}(q^{(i)}) 		& 0          \\
						C_{cl,u}(q^{(i)})Q_i\!\!    &     D_{cl,uq}(q^{(i)}) U +D_{cl,uv}(q^{(i)})X -U        &  D_{cl,uw}(q^{(i)}) 		& 0         \\
								0            &      								0	               &   -I/2 				& 0\\
				C_{cl,z}(q^{(i)})Q_i \! \!      &      D_{cl,zq}(q^{(i)})U +D_{cl,zv}(q^{(i)})X           &    D_{cl,zw}(q^{(i)})   &\!\!\! -\frac{\gamma^2}{2}I
	\end{bmatrix}\! < \! 0\\
&          \begin{bmatrix}
              Q_{i}       &      Y_{i,[k]}^T          \\
	    Y_{i,[k]}            &     \bar{u}_k^2/s^2
	\end{bmatrix} \geq 0, \quad k=1,\ldots, n_u, \quad i=1,\ldots,N
\end{array}
\end{equation}
\begin{center}
\line(1,0){500}
\end{center}
\end{figure*}

In particular, our first anti-windup theorem provides a convex optimization procedure to obtain probabilistic information about the worst case nonlinear \Ltwo\ gain.
To this end, we fix an upper bound $s$ for $\left\|w\right\|_2$ and define two scalars $\epsilon$ and $\delta$ in $(0,1)$ denoting, respectively, an acceptable level of probability of constraint violation and a level of confidence.
Then, inspired by (\ref{eq:L2_computation_analysis}), we apply Theorem~\ref{them:SwC} with the design variables $\theta$ and the certificates $\xi$ given, respectively, by $\theta=\gamma^2$ and $\xi=\left\{ Q, U , Y \right\}$ and the number $N$ of samples selected, based on bound (\ref{eq:binom_reprise}), to satisfy 
\begin{equation}
\B(N,\epsilon,n_{\theta})\leq\delta.
\label{eq:deltabound}
\end{equation}
Then the following result is a straightforward consequence of Theorem~\ref{them:SwC} and Proposition~\ref{prop:AWanalysis}.

\smallskip

\begin{theorem}[Probabilistic performance analysis]
\label{thm:analysis}
Given scalars $s>0$, and $\epsilon,\ \delta \in (0,1)$,
select $N$ satisfying (\ref{eq:deltabound}), fix
$\theta=\gamma^2$ and $\xi=\left\{ Q, U , Y \right\}$.

 If the scenario approximation (\ref{eq:scenario_cert}) of problem \eqref{eq:RwC-AWA} is feasible and attains a unique optimal solution, then for each $\left\|w\right\|_2<s$, the zero initial state solution of system \eqref{eq:CL} satisfies 
\[
\Pr(\left\|z\right\|_2 > \hat{\gamma}(s) \left\|w\right\|_2)<\epsilon,
\] 
with level of confidence no smaller than $1-\delta$.
\end{theorem}

\smallskip

Our second anti-windup theorem allows for robust randomized 
synthesis using the SwC approach and follows parallel steps to those
of Theorem~\ref{thm:analysis} by combining Theorem~\ref{them:SwC} with
Proposition~\ref{prop:AWsynthesis}. To this end, and following
(\ref{eq:L2_computation_synthesis}), we choose the design variables 
$\theta$ and the certificates $\xi$ as follows:
$\theta=\left\{\gamma^2,X,U\right\}$ and $\xi=\left\{ Q, Y
\right\}$. Indeed, the variables $\theta$ must include the quantities $X$
and $U$ used to determine the anti-windup gain in
(\ref{eq:staticAWsel}): these variables must be the same over all
sample extractions so that a unique anti-windup gain can be
determined.
Then the following holds combining Theorem~\ref{them:SwC} with
Proposition~\ref{prop:AWsynthesis}.
 
\smallskip

\begin{theorem}[Probabilistic anti-windup synthesis]
\label{thm:synthesis}
Given scalars $s>0$, and $\epsilon,\ \delta \in (0,1)$,
select $N$ satisfying (\ref{eq:deltabound}), fix
$\theta=\left\{\gamma^2,X,U\right\}$ and $\xi=\left\{ Q, Y
\right\}$.

 If the scenario approximation (\ref{eq:scenario_cert}) of problem \eqref{eq:RwC-AWS} is feasible and attains a unique optimal solution, then for each $\left\|w\right\|_2<s$, the zero initial state solution of the uncertain system \eqref{eq:CL} with anti-windup static compensator \eqref{eq:staticAWsel} satisfies
\[
\Pr(\left\|z\right\|_2 > \hat{\gamma}(s) \left\|w\right\|_2)<\epsilon,
\] 
with level of confidence no smaller than $1-\delta$.
\end{theorem}

For completeness, in \eqref{eq:synthesis-SwC} we report the SwC
problem based on the application of
Theorem~\ref{thm:synthesis}. Similarly, the SwC problem based on the
application of Theorem~\ref{thm:analysis} can be constructed following
the same rationale and it is not reported here due to space limitations.

\begin{remark}
The reformulation of the SwC approach for nonlinear gain analysis and
anti-windup synthesis in Theorems~\ref{thm:analysis} and~\ref{thm:synthesis} is appealing from an engineering viewpoint. As a matter of fact, since the $N$ instances of the system matrices are extracted according to the probability distribution of the uncertainty, this solution provides a view of what \textit{may} happen in most of practical situations. Moreover, we stress that the proposed formulation does not constrain the unknown Lyapunov matrices $Q_{i}$'s to be the same for all the sampled perturbations. Instead, it allows them to vary among different samples. This is possible because the $Q_{i}$'s (as well as the $Y_{i}$'s) are only instrumental for the computation of the robust compensator.
Note that, unlike system matrices, \textit{e.g.}, the $A_{cl}(q^{(i)})$'s, which are uncertain by definition, the certificates are unknown but {\em they are not random variables}. 
\end{remark}

\subsection{Uncertain disturbance energy}
\label{sec:ui}

The previous approach can be further modified by observing that the
design is valid only for a given value of $s$, correspondnig to an
upper bound on the disturbance energy $\|w\|_2$ (see 
the guarantees in Theorems~\ref{thm:analysis} and~\ref{thm:synthesis}). However,
randomization makes it possible to change the perspective of anti-windup
augmentation, by considering also $s$ as an uncertain variable, to
take into account the knowledge of a certain known probability
distribution of the energy of the disturbances acting on the system. 

When considering an uncertain disturbance energy, rather than
minimizing the \Ltwo\ gain at a specific value of $s$, we may consider to minimize the curve over a compact range 
$[\underline s,\overline s]$ of values of $s$, possibly being relevant for the specific distribution.

To this end, we represent
the $\mathcal{L}_2$ gain curve estimate by a polynomial curve over the considered interval, that is,
\begin{equation}
\gamma^2(s)= \sum_{k=0}^{n_{\gamma}}\Gamma_ks^k, \quad s\in
[\underline s, \overline s].
\label{eq:paramGamma}
\end{equation}
Then we may compute an upper bound on the area spanned by the \Ltwo\
gain as follows
\[
\int_{\underline s}^{\overline s}\gamma^2(s) \ ds \leq
\int_{\underline s}^{\overline s}\sum_{k=0}^{n_{\gamma}}\Gamma_ks^k \
ds=\sum_{k=0}^{n_{\gamma}}\Gamma_k\cfrac{\overline s^{k+1}- \underline s^{k+1}
}{k+1}.
\]
This leads to the following RwC problem
\begin{align}
&\min_{\Gamma,U,X} \sum_{k=0}^{n_{\gamma}}\cfrac{\overline s^{k+1}- \underline s^{k+1}
}{k+1}\, \Gamma_k
&\label{eq:RwC-AWS-curve-revisited} \\
&\text{s.t. } \{s, \Gamma,U,X\}\in {\mathcal S}(q,s), \quad\forall
q\in\Q, \forall s\in [\underline s,\overline s],
\nonumber
\end{align}
where we used $\Gamma = [\Gamma_0,\ldots, \Gamma_{n_{\gamma}}]^T$ and, according to parametrization (\ref{eq:paramGamma}),
\[
\begin{array}{l}
{\mathcal S}(q,s) := \bigl\{ \{\Gamma,U,X\}\  | \ \exists 
\{Q,Y\}
\text{ satisfying }  (\ref{eq:analysis1}),
(\ref{eq:analysis2}), (\ref{eq:synthesis})\\
\hspace{3cm} \mbox{ with } \gamma^2 \mbox{ replaced by }
\sum\nolimits\limits_{k=0}^{n_{\gamma}}\Gamma_ks^k  \bigr\}.
\end{array}
\]

Notice that the above design problem is still a convex optimization problem, with the only difference that the cost function is not a single value of $\gamma$ (for a nominal $s$), but a set of values of $\gamma$'s. Namely, we want to minimize the area underlying the \Ltwo\ curve parameterized as a polynomial curve. The problem constraints can still be formulated as LMI's.

The following result is a straightforward generalization of the
construction in Theorem~\ref{thm:synthesis} for this new optimization goal.

\begin{theorem}[Probabilistic synthesis with uncertain energy]
\label{thm:synthesis_ui}
Given scalars $\epsilon,\ \delta \in (0,1)$,
select $N$ satisfying (\ref{eq:deltabound}), fix
$\theta=\left\{\Gamma,X,U\right\}$ and $\xi=\left\{ Q, Y
\right\}$.

 If the scenario approximation (\ref{eq:scenario_cert}) of problem
 \eqref{eq:RwC-AWS-curve-revisited}  is feasible and attains a unique optimal solution,
 then the zero initial state solution of
the uncertain system \eqref{eq:CL} with static anti-windup compensator
\eqref{eq:staticAWsel} satisfies 
\[
\Pr(\left\|z\right\|_2 > \bar{\gamma}(s) \left\|w\right\|_2)<\epsilon,
\] 
with level of confidence no smaller than $1-\delta$, where $\bar
{\gamma}(s) = \sqrt{\gamma^2(s)}$, and $\gamma^2(s)$ defined in (\ref{eq:paramGamma}).
\end{theorem}

\begin{remark}Notice that, when also the input is uncertain, an additional tuning knob appears, namely $n_{\gamma}$, which characterizes the trade-off between computational load and conservativeness. On the one hand, more additional parameters mean a more difficult optimization problem. On the other hand, if the $\mathcal{L}_2$ gain curve is well approximated by the selected polynomial expansion, the upper bound is tight. From practical experience \cite{ZackAWbook11}, the gain curves are typically sigmoidal or exponential functions. Then small values of $n_{\gamma}$ are already enough to obtain good results (usually, from $3$ to $6$). Notice that other basis functions whose integral is linearly parameterized can be suitably selected, without any conceptual change.\end{remark}

\begin{figure*}
%\begin{center}
%\line(1,0){500}
%\end{center}
\begin{subequations}
\label{eq:doa}
\begin{align}
\label{eq:doa1}
    &  Q =Q^T>  0, \; U  > 0 \mbox{ diagonal}, \quad \bar{Q}\leq Q,       \\  
\label{eq:doa2}
       &   \He\begin{bmatrix}
                   A_{cl}(q)Q        &      B_{cl,q}(q)+B_{cl,v}(q)X            \\
	              C_{cl,u}(q)Q-Y    &     D_{cl,uq}(q)U+D_{cl,uv}(q)X-U
	\end{bmatrix} < 0,\\	
&        \label{eq:doa3}      \begin{bmatrix}
              \bar{u}_k^2  &      Y_{[k]}          \\
	           Y_{[k]} ^T    &     Q
	\end{bmatrix} \geq 0, \quad k=1,\ldots, n_u,
\end{align}
\end{subequations}
\begin{center}
\line(1,0){500}
\end{center}
\begin{subequations}
\label{eq:reachableset}
\begin{align}
\label{eq:reachableset1}
    &  Q =Q^T>  0, \; U  > 0 \mbox{ diagonal}, \quad s^2\bar{Q}\geq Q,       \\  
\label{eq:reachableset2}
       &   \He\begin{bmatrix}
                   A_{cl}(q)Q        &      B_{cl,q}(q)+B_{cl,v}(q)X            &   B_{cl,w}(q) \\
	              C_{cl,u}(q)Q-Y    &     D_{cl,uq}(q)U+D_{cl,uv}(q)X-U         &  D_{cl,uw}(q) \\
	                    0                &                                              0               &   -I/2
	\end{bmatrix} < 0,\\	
&        \label{eq:reachableset3}      \begin{bmatrix}
              \bar{u}_k^2/s^2       &      Y_{[k]}          \\
	    Y_{[k]} ^T           &     Q
	\end{bmatrix} \geq 0, \quad k=1,\ldots, n_u,
\end{align}
\end{subequations}
\begin{center}
\line(1,0){500}
\end{center}\end{figure*}

\subsection{Optimized domain of attraction and reachable set}
\label{sec:domain}

Similar derivations to the ones of the previous sections can be obtained by
focusing on different performance goals, as well characterized in
\cite{TarbouriechTAC05} (see also \cite{HuAuto08}). 
In particular, two performance goals which have been well characterized within the context of the use of
generalized sectors for saturated systems correspond to: i) maximizing the
size of a quadratic estimate of the domain of attraction of the origin in the
absence of disturbances (that is, $w=0$), ii) minimizing the best 
quadratic estimate of the reachable set from zero initial conditions
and in the presence of a bounded disturbance $\|w\|_2 \leq s$. 

The goal of this section is then to briefly overview the possible
extensions of the results in Theorems~\ref{thm:synthesis}
and~\ref{thm:synthesis_ui} to these two cases. The following two
propositions establish the baseline results, proven in
\cite{TarbouriechTAC05,HuAuto08} for the nominal case.

\begin{proposition}[Domain of attraction]
\label{prop:doa}
Given the plant-controller pair (\ref{eq:P}), (\ref{eq:C}) and a matrix
$\bar Q = \bar Q^T>0$,
consider the nominal system, that is $\Q\equiv\{\bar q\}$ is a singleton. Assume that the SDP problem 
	\begin{eqnarray}\label{eq:doacomp}
	\displaystyle\max_{\left\{\bar{Q},Q,Y,U,X\right\}} && \log\det(\bar{Q})\\
	&& \mbox{\rm s.t. }(\ref{eq:doa}), \quad \forall q\in \Q,  \nonumber
	\end{eqnarray}
is feasible.
Then, selecting the static anti-windup gain as in
\eqref{eq:staticAWsel}, the nonlinear algebraic loop in (\ref{eq:CL})
is well posed and for any initial condition $x(0)$ in the set
\begin{eqnarray}
{\mathcal E}(\bar{Q}^{-1}) := \{x:\; x^T \bar{Q}^{-1} x \leq 1\},
\label{eq:stab_set}
\end{eqnarray}
the (unique) solution $x$ to the anti-windup closed loop with $w=0$
satisfies $\lim\nolimits\limits_{t\to \infty} |x(t)| = 0$.
\end{proposition}

\smallskip

\begin{proposition}[Reachable set]
\label{prop:reachableset}
Given the plant-controller pair (\ref{eq:P}), (\ref{eq:C}), and a scalar $s>0$,
consider the nominal system, that is $\Q\equiv\{\bar q\}$ is a singleton. Assume that the SDP problem 
	\begin{eqnarray}\label{eq:reachset_comp}
	\displaystyle\min_{\left\{Q,Y,U,X\right\}} && \trace(\bar{Q})\\
	 && \mbox{\rm s.t. }(\ref{eq:reachableset}), \quad \forall q\in \Q, \nonumber
	\end{eqnarray}
is feasible.
Then, selecting the static anti-windup gain as in
\eqref{eq:staticAWsel}, the nonlinear algebraic loop in (\ref{eq:CL})
is well posed and any solution from $x(0)=0$ with
$\left\|w\right\|_2\leq s$ satisfies 
$$
x(t) \in {\mathcal E}(\bar{Q}^{-1}) = \{x:\; x^T \bar{Q}^{-1} x
\leq 1\}, \quad \forall t\geq 0.
$$
\end{proposition}

In light of the results summarized above, we can formulate robust
optimal design and analysis exploiting the constraints
(\ref{eq:doa}) and (\ref{eq:reachableset}), respectively, and leading
to randomized analysis and synthesis tools. These are stated below in
two theorems whose formulations parallel the one of Theorem~\ref{thm:synthesis}.
Analysis results can also be easily stated, paralleling the
formulation in Theorem~\ref{thm:analysis}, but are omitted due to
their straightforward nature, and to avoid
overloading the exposition.

\begin{theorem}[Robust domain of attraction]
\label{thm:robust_doa}
Given scalars $\epsilon,\ \delta \in (0,1)$,
select $N$ satisfying (\ref{eq:deltabound}), fix
$\theta=\left\{\bar Q, X,U\right\}$ and $\xi=\left\{ Q, Y
\right\}$.

If for a selection of $\bar Q = \bar Q^T>0$ and a scalar $\alpha>0$
 the scenario approximation (\ref{eq:scenario_cert}) of problem
 \eqref{eq:doacomp} is feasible and attains a unique optimal solution,
 then for any initial condition in the set (\ref{eq:stab_set}), 
 any solution $x$ of
the uncertain system \eqref{eq:CL} with anti-windup static compensator
\eqref{eq:staticAWsel} and with $w=0$ satisfies for all $t\geq 0$,
\[
x(0) \in {\mathcal E}(\bar{Q}^{-1}) \Rightarrow
\Pr\left(\lim\nolimits\limits_{t\to \infty} |x(t)| = 0 \right) \geq 1-\epsilon,
\] 
with level of confidence no smaller than $1-\delta$.
\end{theorem}

In Theorem~\ref{thm:robust_doa} we characterize properties 
of the scenario approximation (\ref{eq:scenario_cert}) of problem \eqref{eq:doacomp} with the certificates $\xi=\left\{ Q, Y \right\}$. Then, according to the definition in (\ref{eq:S_of_q}), it becomes clear that constraints (\ref{eq:doa}) are imposed with certificates $\left\{ Q, Y \right\}$ depending on the uncertainty $q$, which lead to reduced conservativeness. An interesting feature arising from these $q$-dependent certificates in (\ref{eq:doa}) is that the rightmost constraint in (\ref{eq:doa1}) implies that $\bar Q$ is a uniform lower bound on all certificates $Q_i$. Stated otherwise, this implies that $\mathcal{E}(\bar{Q}^{-1}) \subset \mathcal{E}(Q_i^{-1}), \quad i = 1, \ldots,N$, namely set $\mathcal{E}(\bar{Q}^{-1})$ is a subset of all the stability regions $\mathcal{E}(Q_i^{-1})$ obtained for each one of the extracted samples $q_i$. Then, differently from classical deterministic approaches, although the set $\mathcal{E}(\bar{Q}^{-1})$ is a guaranteed region of robust stability, it is not necessarily a forward invariant set (whereas for each $q_i$ we know that  $\mathcal{E}(Q_i^{-1})$ is a forward invariant set).

A similar (but somewhat converse) comment applies to the robust reachable set studied in the theorem below, wherein the rightmost inequality in \eqref{eq:reachableset1} implies that for each $q_i$ we have $\mathcal{E}(s^2Q_i^{-1})  \subset  \mathcal{E}(\bar{Q}^{-1}), \quad i = 1, \ldots,N$, namely  set $\mathcal{E}(\bar{Q}^{-1})$ is a superset of all the reachable set estimates $\mathcal{E}(s^2Q_i^{-1})$ obtained from the scenario approximation of \eqref{eq:reachset_comp}.

\begin{theorem}[Robust reachable set]
\label{thm:robust_reachableset}
Given scalars $s>0$, and $\epsilon,\ \delta \in (0,1)$,
select $N$ satisfying (\ref{eq:deltabound}), fix
$\theta=\left\{\bar Q, X,U\right\}$ and $\xi=\left\{ Q, Y
\right\}$.

 If the scenario approximation (\ref{eq:scenario_cert}) of problem \eqref{eq:reachset_comp} is feasible and attains a unique optimal solution, then 
for each $\left\|w\right\|_2<s$, the zero initial state solution of
the uncertain system \eqref{eq:CL} with anti-windup static compensator
\eqref{eq:staticAWsel} satisfies 
\[
\Pr\left( x(t) \notin  {\mathcal E}(\bar{Q}^{-1})  \right) <\epsilon
\] 
for all $t\geq 0$, with level of confidence no smaller than $1-\delta$.
\end{theorem}

\begin{remark}
Notice that Theorem~\ref{thm:robust_reachableset} proposes a selection
of the anti-windup gain that minimizes a suitable measure 
of the size of the reachable set for a specific selection of the bound
$s$ on the \Ltwo\ norm of the disturbance $w$. It is then possible to
follow similar derivations to those given in Section~\ref{sec:ui} with
the goal of providing a suitably weighted optimal selection of
the anti-windup gain performed by focusing on the size of the
reachable set in the presence of an unknown \Ltwo\ norm of the
disturbance, for which probabilistic information is available. Then
one may quantify the ``size'' of the reachable set for each value of
$s$ by a suitable parametrization similar to the right hand side of
(\ref{eq:paramGamma}), and finally minimize some net performance
metric taking into account the whole range of possible occurrences of
the \Ltwo\ norm $s$ of the disturbance $w$. Since this extension is
straightforward, it is not discussed in greater detail.
\end{remark}

%%%%%%%%%%%%%%%%%%%%%%%%%%%%%%%%%%%%%%%%%%%%%%%%%%%%%%%%%%%%%%%%%%%%%%%%%%%%%%%%%%%%%%%%%%%%%%%%%%%%%%%%%%%%%%%%%%%%%%%%%%%%
\section{Simulation examples}\label{sec:example}

The following numerical results are obtained using Matlab R2015a on a 64 bit Windows 8.1 computer equipped with an Intel(R) Core(TM) i7-4500U at 1.80 GHz and 8 GB of memory. The optimization is implemented using Yalmip \cite{lofberg2004yalmip} and Sedumi \cite{sturm1999using}.
 
%%%%%%%%%%%%%%%%%%%%%%%%%%%%%%%%%%%%%%%%%%%%%%%%%%%%%%%%
\subsection{$\mathcal{L}_2$ gain minimization}\label{sec:simA}
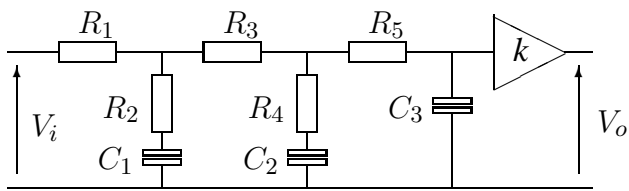
\begin{figure}[hbt]
\begin{center}
\setlength{\unitlength}{1547sp}%
\begingroup\makeatletter\ifx\SetFigFont\undefined%
\gdef\SetFigFont#1#2#3#4#5{%
  \reset@font\fontsize{#1}{#2pt}%
  \fontfamily{#3}\fontseries{#4}\fontshape{#5}%
  \selectfont}%
\fi\endgroup%
\begin{picture}(9687,2799)(1864,-4348)
\thinlines
\put(1876,-2161){\line( 1, 0){825}}
\put(2701,-2311){\framebox(900,300){}}
\put(3601,-2161){\line( 1, 0){1425}}
\put(4351,-2161){\line( 0,-1){375}}
\put(4201,-3436){\framebox(300,900){}}
\put(4351,-3436){\line( 0,-1){300}}
\put(4051,-3811){\framebox(600,75){}}
\put(4051,-3961){\framebox(600,75){}}
\put(4351,-3961){\line( 0,-1){375}}
\put(5926,-2161){\line( 1, 0){1425}}
\put(6676,-2161){\line( 0,-1){375}}
\put(6526,-3436){\framebox(300,900){}}
\put(6676,-3436){\line( 0,-1){300}}
\put(6376,-3811){\framebox(600,75){}}
\put(6376,-3961){\framebox(600,75){}}
\put(6676,-3961){\line( 0,-1){375}}
\put(5026,-2311){\framebox(900,300){}}
\put(7351,-2311){\framebox(900,300){}}
\put(8251,-2161){\line( 1, 0){1425}}
\put(9001,-2161){\line( 0,-1){750}}
\put(8701,-2986){\framebox(600,75){}}
\put(8701,-3136){\framebox(600,75){}}
\put(9001,-3136){\line( 0,-1){1200}}
\put(9676,-1561){\line( 0,-1){1200}}
\put(9676,-2761){\line( 2, 1){1140}}
\put(10801,-2161){\line(-2, 1){1140}}
\put(10801,-2161){\line( 1, 0){450}}
\put(1876,-4336){\line( 1, 0){9375}}
\put(2026,-4111){\vector( 0, 1){1725}}
\put(11026,-4111){\vector( 0, 1){1725}}
\put(2251,-3511){\makebox(0,0)[lb]{\smash{\SetFigFont{13}{34.8}{\rmdefault}{\mddefault}{\itdefault}$V_{i}$}}}
\put(11326,-3436){\makebox(0,0)[lb]{\smash{\SetFigFont{13}{34.8}{\rmdefault}{\mddefault}{\itdefault}$V_{o}$}}}
\put(5651,-4036){\makebox(0,0)[lb]{\smash{\SetFigFont{13}{34.8}{\rmdefault}{\mddefault}{\itdefault}$C_{2}$}}}
\put(7976,-3211){\makebox(0,0)[lb]{\smash{\SetFigFont{13}{34.8}{\rmdefault}{\mddefault}{\itdefault}$C_{3}$}}}
\put(3326,-4036){\makebox(0,0)[lb]{\smash{\SetFigFont{13}{34.8}{\rmdefault}{\mddefault}{\itdefault}$C_{1}$}}}
\put(3401,-3211){\makebox(0,0)[lb]{\smash{\SetFigFont{13}{34.8}{\rmdefault}{\mddefault}{\itdefault}$R_{2}$}}}
\put(5726,-3211){\makebox(0,0)[lb]{\smash{\SetFigFont{13}{34.8}{\rmdefault}{\mddefault}{\itdefault}$R_{4}$}}}
\put(9976,-2311){\makebox(0,0)[lb]{\smash{\SetFigFont{13}{34.8}{\rmdefault}{\mddefault}{\itdefault}k}}}
\put(7651,-1861){\makebox(0,0)[lb]{\smash{\SetFigFont{13}{34.8}{\rmdefault}{\mddefault}{\itdefault}$R_{5}$}}}
\put(5326,-1861){\makebox(0,0)[lb]{\smash{\SetFigFont{13}{34.8}{\rmdefault}{\mddefault}{\itdefault}$R_{3}$}}}
\put(3001,-1861){\makebox(0,0)[lb]{\smash{\SetFigFont{13}{34.8}{\rmdefault}{\mddefault}{\itdefault}$R_{1}$}}}
\end{picture}
\caption{\label{fig:network} The passive electrical network with saturated input voltage.}
\end{center}
\end{figure}

In this section, we show the effectiveness of the proposed approach, by designing an anti-windup compensator for the passive electrical network in Fig.~\ref{fig:network}. The circuit is a benchmark example in the anti-windup literature and was already employed in \cite{ZackAWbook11} to show the potential of static anti-windup in a deterministic context.

%%%%%%%%%%%%%%%%%%%%%%%%%%%%%%%%%%%%%%%%%%%%%%%%%%%%%%%%%%%%
The dynamics of the network is determined by 5 resistors and 3 capacitors, whose nominal values are reported in Table~\ref{tab:parameters}. The gain $k$ is instead selected such that the transfer function between $V_i$ and $V_o$ is monic.

After some cumbersome computations, the transfer function of the network turns out to be
\begin{equation}
G(s) =  \frac{s^2 + \frac{C_1 R_2+C_2 R_4}{C_1 C_2 R_2 R_4}s + \frac{1}{C_1 C_2 R_2 R_4}}{s^3 + \frac{\eta_2}{\eta_3} s^2+ \frac{\eta_1}{\eta_3} s + \frac{1}{\eta_3}}
\end{equation}
with
\begin{align}
\eta_1 &= C_1 R_1 + C_1 R_2 + C_2 R_3 + C_2 R_4 + C_3 R_5,\\
\eta_2 &= C_1 C_2 R_1 R_3 + C_1 C_2 R_1 R_4 + C_1 C_2 R_2 R_3\nonumber\\
       & + C_1 C_2 R_2 R_4 + C_1 C_3 R_1 R_5 + C_1 C_3 R_2 R_5\nonumber\\
       &+ C_2 C_3 R_3 R_5 + C_2 C_3 R_4 R_5,\\
\eta_3 &= C_1 C_2 C_3 R_1 R_3 R_5 + C_1 C_2 C_3 R_1 R_4 R_5\nonumber\\
       &+ C_1 C_2 C_3 R_2 R_3 R_5 + C_1 C_2 C_3 R_2 R_4 R_5.
\end{align}
Notice that the dependence of $\eta_i$, $i=1,2,3$ upon the physical parameters is highly nonlinear.

%%%%%%% PARAMETERS %%%%%%%%%%%%%%%%%%%%%%%%%%%%%%%%%%%%%%%%%
\begin{table}
\centering
\begin{tabular}{c||c|c}
\textbf{name}&\textbf{value}&\textbf{units}\\
\hline
$R_1$ & $313$  &$\mathrm{\Omega}$\\ 
$R_2$ & $20$   &$\mathrm{\Omega}$\\ 
$R_3$ & $315$  &$\mathrm{\Omega}$\\ 
$R_4$ & $17$   &$\mathrm{\Omega}$\\
$R_5$ & $10$   &$\mathrm{\Omega}$\\ 
$C_1$ & $0.01$ &$\mathrm{F}$\\
$C_2$ & $0.01$ &$\mathrm{F}$\\ 
$C_3$ & $0.01$ &$\mathrm{F}$\\
\end{tabular}
\vspace{3mm}
\caption{Nominal parameter values for the network in Fig.~\ref{fig:network}.}
\label{tab:parameters}
\end{table}
%%%%%%%%%%%%%%%%%%%%%%%%%%%%%%%%%%%%%%%%%%%%%%%%%%%%%%%%%%%%

The nominal plant can then be put in the form \eqref{eq:P} via suitable state-space realization, where
\begin{align}
\footnotesize
\left[
\begin{array}{c|c|c}
A_p     & B_{p,u}   & B_{p,w}\\    \hline
C_{p,z} & D_{p,zu}  &   D_{p,zw}\\ \hline
C_{p,y} & D_{p,yu}  &   D_{p,yw}
\end{array}
\right] =
\left[
\begin{array}{ccc|c|c}
 -10.6  &  -6.09  & -0.9 & 1 &0\\ 
    1   &    0    &   0  & 0 &0\\
    0   &    1    &   0  & 0 &0\\ \hline
   -1   &   -11   &  -30 & 0 &0\\ \hline
    1   &   11    &  30 & 0 &0\\
\end{array}
\right], \nonumber
\normalsize
\end{align}
and $w$ represents the reference value for the output voltage $V_o$, so that $z=w-y$ is the tracking error.

\begin{figure}[h!]
\centering
\includegraphics[width = 1\columnwidth]{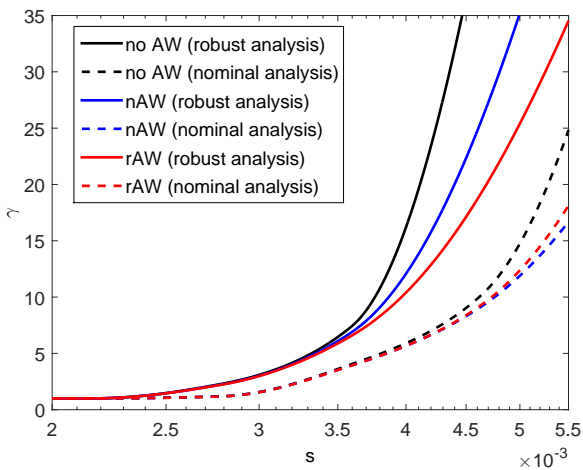}
\caption{\Ltwo\ gain estimates for the nonlinear closed-loop systems with and without anti-windup compensator. Both robust (solid) and nominal (dashed) analysis are considered to assess the performance of robust (blue, $D_{\rm{aw}}^{\rm{rand}}$) and nominal (red, $D_{\rm{aw}}^{\rm{nom}}$) compensators with respect to the system without anti-windup augmentation (black).}
\label{fig:L2gain}
\end{figure}

The controller is a PID and it is designed based on the nominal model, such that the nominal phase margin is $89.5$ degrees and the nominal gain margin is infinity. In the form \eqref{eq:C}, the controller is expressed by the matrices
\begin{align}
&\left[
\begin{array}{c|c|c}
A_c   & B_{c,y} & B_{c,w} \\ \hline
C_{c} & D_{c,y} & D_{c,w}
\end{array}
\right] =
\left[
\begin{array}{cc|c|c}
-80   & 0    & 1  &-1\\ 
1     & 0    & 0  &0\\ \hline
20.25 & 1600 & 80 &-80\\ 
\end{array}
\right].\nonumber
\end{align}
Assume now that the input $V_i$ is saturated between the minimum and the maximum voltages $\pm 1$. For a specific value of $s$, a static anti-windup compensator based on the nominal model can be designed to minimize the nonlinear gain between the reference and the tracking error, by solving the optimization problem \eqref{eq:L2_computation_synthesis} and relying on Proposition~\ref{prop:AWsynthesis}. Using $s=0.003$, the \emph{nominal} optimal value $\hat \gamma_n^2(s)=2.31$ is obtained, together with the optimal nominal anti-windup gain $D^{\rm{nom}}_{\rm{aw}}=[-0.0855,0.0011,0.9887]^T$. This value corresponds to the squared value of the dashed blue curve in Fig.~\ref{fig:L2gain} at the abscissa $s=0.003\approx 10^{-2.52}$. 

Under the hypothesis that the parameters are Gaussian distributed with mean values as in Table I and standard deviation of $10\%$, a robust randomized compensator can be computed following Theorem~\ref{thm:synthesis}. First notice that $n_{\theta}=5 =1+3+1$ (arising from $\gamma^2$, $X$ and $U$, respectively). Then, fixing parameters $\varepsilon=0.01$ and $\delta=10^{-6}$, we see that $N=2819$ samples are necessary to satisfy \eqref{eq:N}. Therefore, we follow the sequential algorithm of Section~\ref{sec:SwC} to reduce the computational burden. In particular, using the same value $s=0.003$ as in the nominal synthesis above, we apply the Sequential algorithm for SwC, initialized with $k_t=10$. Such a procedure terminates after $3$ iterations, using only $N = 846$ samples, and providing the robust optimal value $\hat \gamma_r^2=9.1$ (evidently larger than the nominal one), together with the optimal robust anti-windup gain $D^{\rm{rand}}_{\rm{aw}}=[-2.1493,0.0266,0.6407]^T$. This value approximately corresponds to the squared value of the solid red curve in Fig.~\ref{fig:L2gain} at the abscissa $s=0.003\approx 10^{-2.52}$. We observe a slight difference between the two values (in the figure, the gain is higher), justified by the fact that the performance analysis is carried out with a different set of samples. 
We should remark that in terms of computational time, the SwC approach is more demanding than the nominal design. In this example, the elapsed time for compensator design is approximately $7$ seconds in the latter case and about $5$ hours in the former.

%%%%%%% L2 GAIN %%%%%%%%%%%%%%%%%%%%%%%%%%%%%%%%%%%%%%%%%%%%
%%%%%%%%%%%%%%%%%%%%%%%%%%%%%%%%%%%%%%%%%%%%%%%%%%%%%%%%%%%%
Once the nominal and the robust anti-windup gains are fixed, we may characterize their nominal and robust performance by applying, respectively, the analysis tools of Proposition~\ref{prop:AWanalysis} and Theorem~\ref{thm:analysis}. For comparison purposes, we study the nominal and robust performance also for the case with no anti-windup compensation. Comprehensively, we obtain six curves, all reported in Fig.~\ref{fig:L2gain}, where the nominal curves are dashed and the robust ones are solid. 

As expected, we observe that the robust compensator outperforms the one designed for the nominal system, as far as the robust \Ltwo\ gain is concerned (solid curves). Conversely, when the performance is evaluated on the nominal system (dashed curves), the robust compensator yields worse results, since it is more conservative. From the analysis point of view, notice also that the \Ltwo\ gains estimated using the robust probabilistic method are larger than the ones given by the nominal analysis. This holds for any configuration of the saturated closed-loop system (without anti-windup, with nominal compensator and with robust compensator).

\begin{figure}[h!]
\centering
\includegraphics[width = 1\columnwidth]{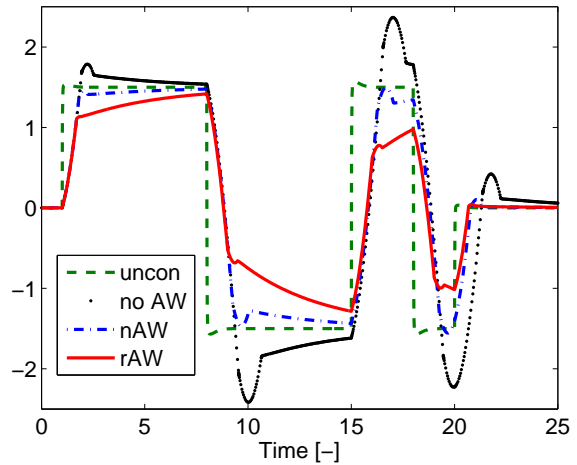}
\caption{Time responses of the closed-loop system with nominal parameters and different configurations of the anti-windup architecture: unconstrained system (dashed), saturated system without anti-windup compensator (dotted), saturated system with nominal anti-windup $D_{\rm{aw}}^{\rm{nom}}$ (dash-dotted) and saturated system with robust anti-windup $D_{\rm{aw}}^{\rm{rand}}$ (solid).}
\label{fig:time_nominal}
\end{figure}

\begin{figure}[h!]
\centering
\includegraphics[width = 1\columnwidth]{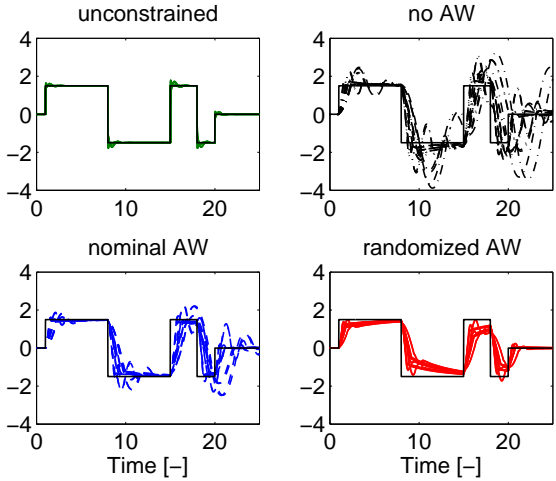}
\caption{Sixteen perturbed time responses of the uncertain closed-loop system with different configurations of the anti-windup architecture: unconstrained system (upper-left), saturated system without anti-windup compensator (upper-right), saturated system with nominal anti-windup $D_{\rm{aw}}^{\rm{nom}}$ (lower left) and saturated system with robust anti-windup $D_{\rm{aw}}^{\rm{rand}}$ (lower right).}
\label{fig:time_robust}
\end{figure}

The time-domain performance degradation of the nominal closed-loop system using the robust compensator in place of the nominal one can be assessed by looking at the time responses illustrated in Fig.~\ref{fig:time_nominal}. From the figure, we conclude that, although in any case the use of a compensator (red solid and blue dash-dotted curves) improves upon the response without anti-windup (black dotted) in terms of tracking error and overshoot, we have to accept worse behavior in nominal conditions when using robust anti-windup (indeed, the blue dash-dotted response yields faster transients than the red solid curves).
%%%%%%% TIME NOMINAL %%%%%%%%%%%%%%%%%%%%%%%%%%%%%%%%%%%%%%%
%%%%%%%%%%%%%%%%%%%%%%%%%%%%%%%%%%%%%%%%%%%%%%%%%%%%%%%%%%%%
However, this choice is rewarding when acting on a system subject to uncertainty. Specifically, in Fig.~\ref{fig:time_robust}, we show that the response with the perturbed plant using robust anti-windup is less sensitive to parameter uncertainties as compared to nominal anti-windup. As an example, the figure shows the time responses corresponding to $16$ different combinations of $\pm 10 \%$ perturbations of the nominal parameters.

%%%%%%% TIME ROBUST %%%%%%%%%%%%%%%%%%%%%%%%%%%%%%%%%%%%%%%%
%%%%%%%%%%%%%%%%%%%%%%%%%%%%%%%%%%%%%%%%%%%%%%%%%%%%%%%

It should be here remarked that, if the approach in \cite{FormentinCDC13} is employed, the optimization problem for the design of the anti-windup compensator becomes infeasible. This is due to the conservativeness of the formulation in \cite{FormentinCDC13}, which requires a single Lyapunov function for all the possible uncertain instances of the system.

\subsection{Uncertain disturbance energy}

\begin{figure}[h!]
\centering
\includegraphics[width = 1\columnwidth]{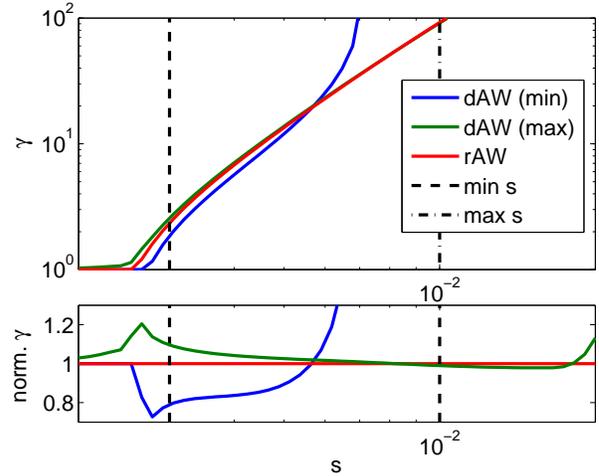}
\caption{Randomized minimization of the $\mathcal{L}_2$ gain considering $s=0.003$ (blue solid, $D^{\rm{nom,min}}_{\rm{aw}}$) and $s=0.01$ (green solid, $D^{\rm{nom,max}}_{\rm{aw}}$), against randomized minimization of the area underlying the curve (red solid, $D^{\rm{rand}}_{\rm{aw}}$).}
\label{fig:area}
\end{figure}

We use the example of the previous section to illustrate the synthesis of Section~\ref{sec:ui} aimed to (probabilistically) minimizing the area spanned by the nonlinear \Ltwo\ gain curve within a given interval. Specifically, we consider the case where the system parameters are fixed, but $s$ is unknown and has a uniform distribution between $\underline{s} = 0.003$ and $\overline{s}=0.01$. To this end, we use a cubic upper bound for the curve (namely, we fix $n_{\gamma}=3$ in \eqref{eq:RwC-AWS-curve-revisited}). Notice that now $n_{\theta}=8$, because $\gamma^2$ is no longer needed, but the parameters $\Gamma_k$, $k=0,\ldots,3$ in \eqref{eq:paramGamma} have to be designed. In this way, the scenario bound \eqref{eq:N} raises up to $N=3293$. However, the sequential algorithm of Section~\ref{sec:SwC} with $k_t=10$ allows us to find the desired solution with ``only" $2306$ samples. The corresponding anti-windup compensator reads $D^{\rm{rand}}_{\rm{aw}}=[-0.0003554,0.0000021,0.9987978]^T$.
The nonlinear \Ltwo\ gain of the closed-loop system with such a compensator corresponds to the red curve represented in Fig.~\ref{fig:area}. To better assess the performance with this anti-windup gain, we also show the nonlinear \Ltwo\ gain curves obtained from the deterministic design of Proposition~\ref{prop:AWsynthesis} corresponding to the minimum $s=\underline{s}=0.003$ (blue curve) and the maximum $\overline{s}=0.01$ (green curve). The corresponding compensators are, respectively, $D^{\rm{nom,min}}_{\rm{aw}}=[-0.022124,0.000276,0.999904]^T$ and $D^{\rm{nom,max}}_{\rm{aw}}=[-0.073639,0.000920,0.999999]^T$. In the lower subplot of Fig.~\ref{fig:area}, the same 
nonlinear gains are normalized in terms of percentage of the red curve. 

%%%%%%% AREA %%%%%%%%%%%%%%%%%%%%%%%%%%%%%%%%%%%%%%%%%%%%
%%%%%%%%%%%%%%%%%%%%%%%%%%%%%%%%%%%%%%%%%%%%%%%%%%%%%%%

Fig.~\ref{fig:area} shows that the minimization of the area underlying the curve may be a good trade-off. 
Indeed, while at $\underline{s}$ the blue curve provides a smaller gain, that blue curve blows up to infinity even before reaching $s=\overline{s}$, thereby providing an optimal behavior at  $\underline{s}$ and not even guaranteeing stability at $\overline{s}$. Similarly, the green curve provides desirable optimal performance at $\overline {s}$ but sacrifices the performance at $\underline{s}$. The red curve clearly shows a trade-off that somewhat sits in the middle between the two extreme green and blue solutions.
Such a result is obtained at the price of some additional computational cost, in that more free parameters are involved (recall the parameterization of the curve in \eqref{eq:paramGamma}). The elapsed time for designing such a compensator is then about $6$ hours (with an increase of $20\%$ with respect to the SwC design of the previous section).

\subsection{Reachable set and domain of attraction}

In this section, we illustrate the effectiveness of the proposed randomized anti-windup design approach when the control objective is the minimization of the reachable set or the maximization of the domain of attraction (see Section~\ref{sec:domain}). To this aim, a simple example with a planar closed loop is considered to easily visualize the obtained sets in proper phase planes.

Consider the first-order plant
\begin{align}
\left[
\begin{array}{c|c|c}
A_p     & B_{p,u}   & B_{p,w}\\    \hline
C_{p,z} & D_{p,zu}  &   D_{p,zw}\\ \hline
C_{p,y} & D_{p,yu}  &   D_{p,yw}
\end{array}
\right] =
\left[
\begin{array}{c|c|c}
 a &  b  & 0\\  \hline
-1 &  0  & 1\\ \hline
 1 &  0  & 0
\end{array}
\right], \nonumber
\end{align}
where $a$ and $b$ are stochastic variables with Gaussian distribution. Specifically, let $\mathbb{E}[a]=-1$, $\mathbb{E}[b]=1$ and their standard deviation be of $20\%$. 

%%%%%%% RS %%%%%%%%%%%%%%%%%%%%%%%%%%%%%%%%%%%%%%%%%%%%
\begin{figure*}	
 \centering
\subfigure[]
   {\includegraphics[width = 1\columnwidth]{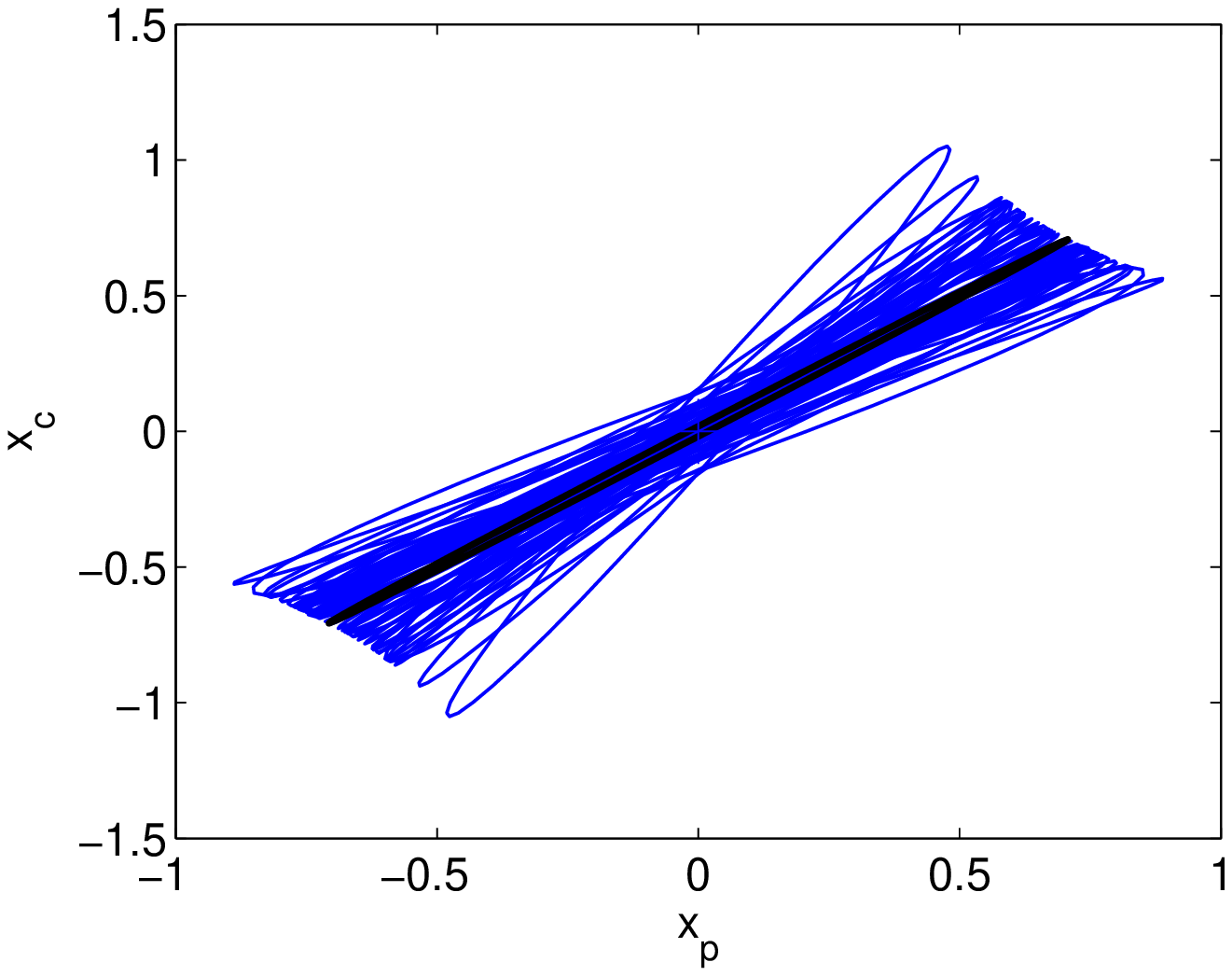}\label{fig:rs_det}}
\subfigure[]
   {\includegraphics[width = 1\columnwidth]{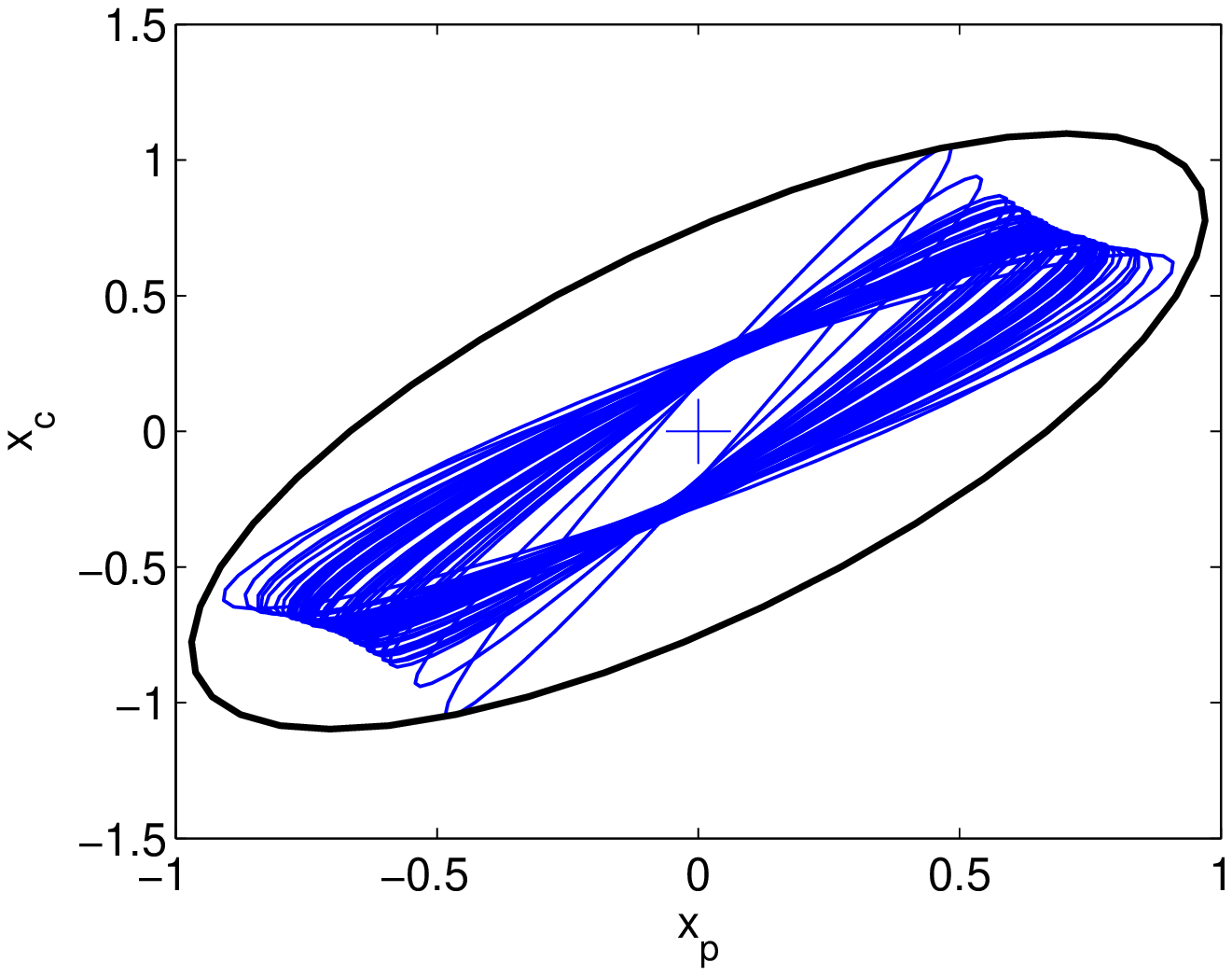}\label{fig:rs_swc}}
   \caption{Reachable sets for the closed-loop system with $50$ different uncertainty samples (blue lines) and estimates of the reachable set given by the deterministic design (black line, left, $D^{\rm{nom}}_{\rm{aw}}$) and SwC design (black line, right, $D^{\rm{rand}}_{\rm{aw}}$).}
   \label{fig:rs}
\end{figure*}

%%%%%%%%%%%%%%%%%%%%%%%%%%%%%%%%%%%%%%%%%%%%%%%%%%%%%%%%

%%%%%%% DOA %%%%%%%%%%%%%%%%%%%%%%%%%%%%%%%%%%%%%%%%%%%%
\begin{figure*}
 \centering
\subfigure[]
   {\includegraphics[width = 1\columnwidth]{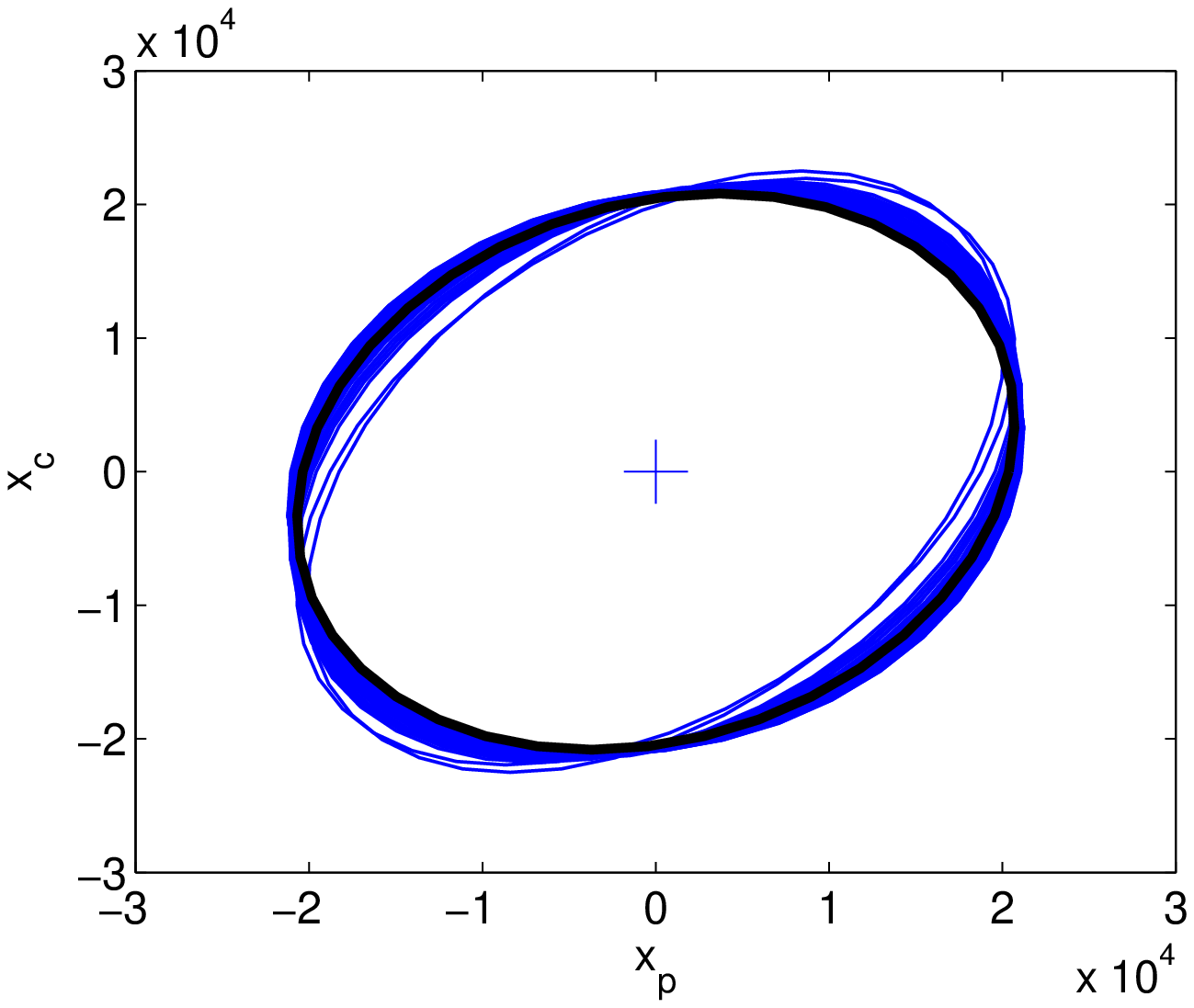}\label{fig:doa_det}}
\subfigure[]
   {\includegraphics[width = 1\columnwidth]{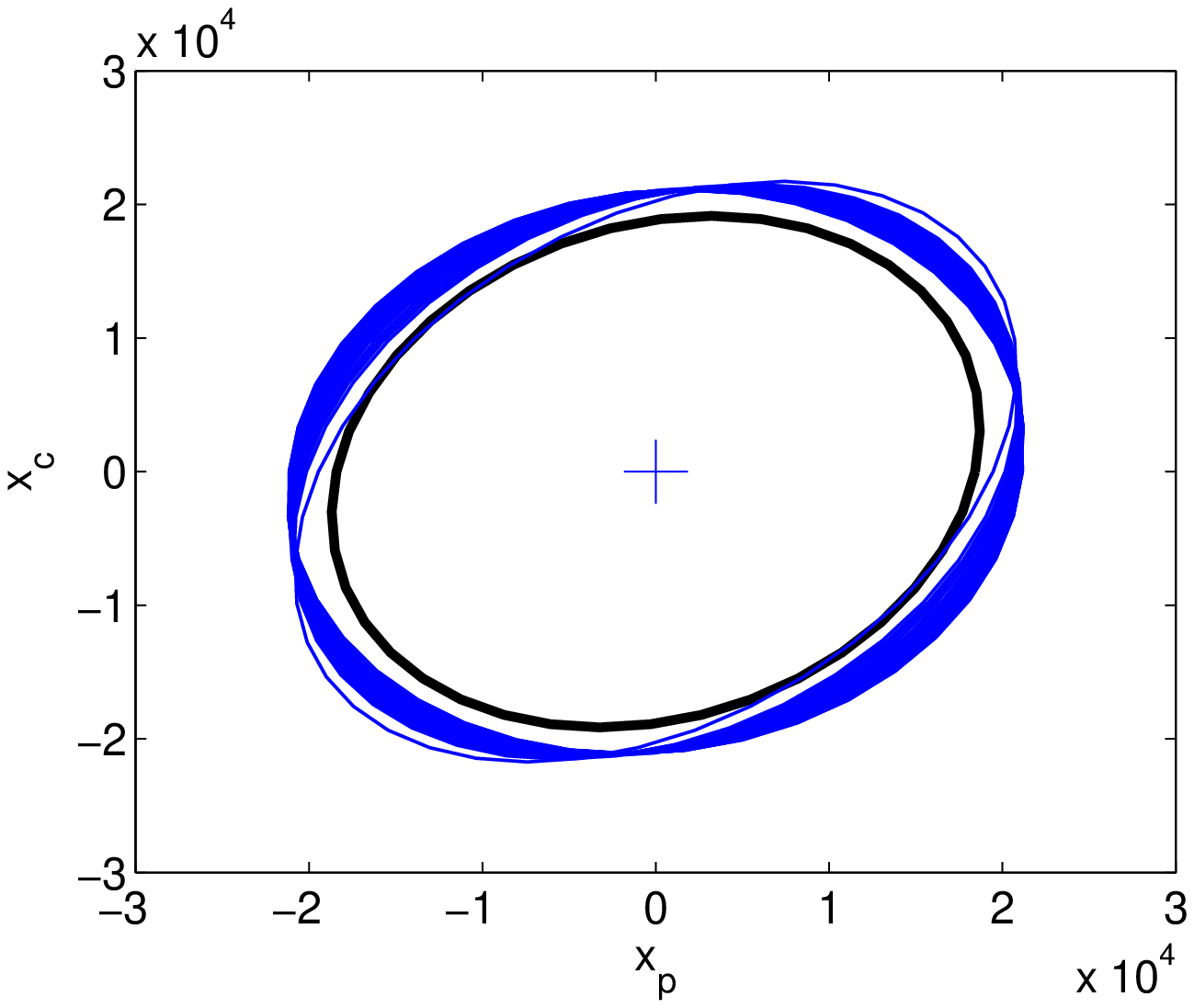}\label{fig:doa_swc}}
   \caption{Domain of attraction for the closed-loop system with $50$ different uncertainty samples (blue lines) and estimates of the domains given by the deterministic design (black line, left, $D^{\rm{nom}}_{\rm{aw}}$) and SwC design (black line, right, $D^{\rm{rand}}_{\rm{aw}}$).}
   \label{fig:doa}
\end{figure*}
%%%%%%%%%%%%%%%%%%%%%%%%%%%%%%%%%%%%%%%%%%%%%%%%%%%%%%%%

The integral controller with unitary gain
\begin{align}
&\left[
\begin{array}{c|c|c}
A_c   & B_{c,y} & B_{c,w} \\ \hline
C_{c} & D_{c,y} & D_{c,w}
\end{array}
\right] =
\left[
\begin{array}{c|c|c}
0   & -1 & 1\\ \hline 
1   & -1 & 1\\ 
\end{array}
\right]\nonumber
\end{align}
is employed, so that the stability of the nominal unconstrained closed-loop system is guaranteed. Suppose now that the input $u$ is bounded below and above by $\underline{u}=-1$ and $\bar{u}=1$, respectively. For both the following probabilistic design procedures, we set $\epsilon=0.01$ and $\delta=10^{-6}$. Since $n_{\theta}=6$, the resulting scenario bound, according to \eqref{eq:N}, turns out to be $N=2977$. The sequential algorithm of Section~\ref{sec:SwC} is run with $k_t=10$.

A robust anti-windup compensator minimizing the reachable set can be designed according to Theorem~\ref{thm:robust_reachableset}. The resulting anti-windup compensator is $D^{\rm{rand}}_{\rm{aw}}=[-0.0212,0.9902]^T$ and is obtained with $N=596$ samples instead of $N=2977$, thanks to the sequential algorithm. Fig.~\ref{fig:rs_swc} shows that the reachable set obtained with such an approach (black thick line) can be considered as an upper bound for the reachable sets of the uncertain closed-loop system with $50$ random samples of the uncertain parameters (blue thin lines). It should be remarked that, being the samples drawn independently from the samples used for design, one reachable set actually crosses the black line (the robust design gives only \emph{probabilistic} certificates). A different conclusion can be drawn when the compensator is designed based on the nominal model only, according to Proposition~\ref{prop:reachableset}, as illustrated in Fig.~\ref{fig:rs_det}. In this case, the anti-windup compensator reads $D^{\rm{nom}}_{\rm{aw}}=[-0.00001,0.99998]^T$, which evidently does not guarantee that the reachable sets achievable with different uncertainty samples (blue thin lines) are included in the estimated set (black thick line). Notice that the samples used to plot the $50$ instances of the uncertain system in Fig.~\ref{fig:rs_det} and Fig.~\ref{fig:rs_swc} are the same.

The same comparison, with analogous conclusions, can be made when the goal of the anti-windup design is the maximization of the domain of attraction. In particular, Proposition~\ref{prop:doa} and Theorem~\ref{thm:robust_doa} yield, respectively, $D^{\rm{nom}}_{\rm{aw}}=[-2.3475,-1.0063]^T$ and $D^{\rm{rand}}_{\rm{aw}}=[-3.2036,-0.9998]^T$, using $N=1191$ samples with the sequential algorithm.

Fig.~\ref{fig:doa_det} clearly shows - also in this case - the limits of the deterministic approach, which takes into account only the nominal values of the system parameters, thus leading to an unreliable estimate of the minimum domain of attraction (black thick line), crossed by many of the uncertain domain samples (blue thin lines). Conversely, in Fig.~\ref{fig:doa_swc}, the randomized compensator guarantees that most (in probability) of the uncertain domains (blue thin lines) contain the estimate of the domain of attraction obtained using the randomized approach (black solid line).

For such a simple example, the elapsed time to compute the robust compensator (for both the domain of attraction and the reachable set) is smaller than in the electrical network example and is about $53$ minutes. This is still significant if compared to the amount of time required for the simple compensator, which is approximately $5$ seconds, but on the other hand, such a tuning provides robustness guarantees otherwise unobtainable with a classical deterministic approach. Moreover, it should be recalled here that the design time has no effect of the on-line computational time for the anti-windup compensation, since the robust compensator is characterized by the same structure of the nominal one.

%%%%%%%%%%%%%%%%%%%%%%%%%%%%%%%%%%%%%%%%%%%%%%%%%%%%%%%%%%%%%%%%%%%%%%%%%%%%%%%%%%%%%%%%%%%%%%%%%%%%%%%%%%%%%%%%%%%%%%%%%%%%
\section{Conclusions}
\label{sec:conclusions}

In this paper we proposed a novel paradigm for approaching static linear anti-windup design for linear saturated control systems in the presence of nonlinear probabilistic uncertainties.
The proposed paradigm relies on randomized approaches and provides a successful tool to tackle this challenging robust analysis and design problem. The peculiar structure of static linear anti-windup design is particularly suited as a promising extension of the typical scenario approach to randomized design. In particular, since the design variable is decoupled from the Lyapunov certificate, we introduce a so-called ``scenario with certificates'' paradigm to provide a dramatically reduced conservatism, as compared to typical approaches, based on common quadratic certificates. 
The randomized approach to anti-windup design may be formulated using a wide range of optimality goals that we study in this paper and for which we illustrate the advantages by way of numerical results.

\section*{Acknowledgments}
This work was supported in part by the ANR project LimICoS contract number 12 BS03 005 01, by CNR funds of the joint international lab COOPS, by iCODE institute, research project of the Idex Paris-Saclay and and by grant OptHySYS funded by the University of Trento.

\ifCLASSOPTIONcaptionsoff
  \newpage
\fi

\bibliographystyle{plain}
\bibliography{randomized_methods,saturate,aw,aw2,zack,Frugi-biblio}

\end{document}